\documentclass[11pt]{article}
\usepackage{latexsym,amsmath,amscd,amssymb, graphics}
\usepackage{amsmath,amssymb,mathrsfs,framed,esint,slashed,color}
\usepackage{amsthm}
\usepackage{enumerate}
\usepackage{hyperref}
\hypersetup{
colorlinks = true, 
urlcolor = blue, 
linkcolor = blue, 
citecolor = red 
}
\usepackage{enumitem}
\usepackage[margin=1in]{geometry}
\usepackage{stmaryrd}
\usepackage{graphicx}
\usepackage[all]{xy}
\usepackage[makeroom]{cancel}
\usepackage{empheq}

\usepackage{framed}

\usepackage{pgfplots, pgfplotstable, booktabs, colortbl, array,multirow}
\usetikzlibrary{calc}
\usepgfplotslibrary{groupplots}
\pgfplotsset{compat=newest}

\theoremstyle{plain}
\newtheorem{theorem}{Theorem}[section]
\newtheorem{lemma}[theorem]{Lemma}

\newtheorem{varia}{Variational principle}

\theoremstyle{definition}

\newtheorem{remark}{Remark}[section]

\usepackage{verbatim}

\begin{document}

\title{Variational formulation of a general dissipative fluid system with differential forms}

\author{
Bastien Manach-P\'erennou\thanks{\noindent Division of Mathematical Sciences, Nanyang Technological University, Singapore
\href{bastien.manachp@ntu.edu.sg}{bastien.manachp@ntu.edu.sg}}
\and
Fran\c{c}ois Gay-Balmaz\thanks{\noindent Division of Mathematical Sciences, Nanyang Technological University, Singapore \href{francois.gb@ntu.edu.sg}{francois.gb@ntu.edu.sg}}
}

\date{}

\maketitle

\begin{abstract}
This work is devoted to the study of dissipative fluid systems, through the lens of a geometric variational formulation. Building upon previous works extending Hamilton's principle to non-equilibrium thermodynamics, the present method incorporates an arbitrary number of additional variables expressed as differential forms. Dissipation sources, thermodynamic flux closures, and their associated boundary conditions are also all expressed in this differential-form framework. The resulting equations are consistent with the fundamental laws of thermodynamics, namely conservation of total energy and positive entropy production. Onsager's principle is also given a simple formulation, while Curie’s principle is revisited within this geometric setting through the lens of representation theory. It is shown that this general framework encompasses physically relevant models, such as multi-species magnetohydrodynamics (MHD) equations with intricate dissipation mechanisms.
\end{abstract}

\section{Introduction}

In this article, we present a new geometric formulation of a variational principle for dissipative fluid systems, emphasising the intrinsic nature of the relevant variables. This approach (i) is based on a variational principle, (ii) is expressed using the geometric language of differential forms, and (iii) extends naturally to non-equilibrium thermodynamics. To illustrate these three aspects and their relevance, we begin with the well-studied example of magnetohydrodynamics (MHD) with viscosity, resistivity, and heat transfer. On a domain $\Omega\subset\mathbb{R}^3$ with smooth boundary and with respect to cartesian coordinates, the governing equations are
\begin{subequations}
\label{eq:MHD}
\begin{align}
\rho (\partial_t u + u \cdot \nabla u ) + \nabla p & = \frac{1}{\mu_0} \left( \nabla \times B \right) \times B + \nabla \cdot \sigma, \\
\partial_t \rho + \nabla \cdot (\rho u) & = 0, \\
\partial_t B + \nabla \times (B \times u) & = \frac{1}{\mu_0} \nabla \times ( \nu \nabla \times B), \\
T\left( \partial_t \mathsf{s} + \nabla \cdot (\mathsf{s} u) - \nabla\cdot \left( \frac{\kappa}{T} \nabla T \right)\right) & = \frac{\kappa}{T} (\nabla T) \cdot (\nabla T) + \frac{\nu}{\mu_0^2} (\nabla \times B) \cdot (\nabla \times B).
\end{align}
\end{subequations}
When the system is assumed isolated, these equations are supplemented with the boundary conditions $B \cdot n=0$, $u=0$, $\kappa \nabla T \cdot n = 0$ and $\frac{\nu}{\mu_0} (\nabla \times B) \times n = 0$ on $\partial\Omega$. All notations are standard: $u$ is the fluid velocity, $\rho$ the mass density, $\mathsf{s}$ the entropy density, and $B$ the magnetic induction. The pressure $p$ and temperature $T$ are prescribed thermodynamic functions of $\rho$ and $\mathsf{s}$. The tensor $\sigma$ denotes the viscous stress. Finally, $\nu$ is the resistivity coefficient, $\kappa T$ the heat conductivity, and $\mu_0$ the vacuum permeability.

\paragraph{Variational principles in the dissipationless case.}
Setting $\nu$, $\kappa$ and $\sigma$ to zero (i.e. neglecting dissipation), the previous equations follow from Hamilton's principle formulated in Lagrangian variables \cite{Hattori1994, Ono1995}. The fluid motion is described by the trajectory map, assumed to be a diffeomorphism of $\Omega$ at all times. When expressed in Eulerian variables, this leads to the variational principle
\begin{equation}
\label{eq:vp_MHD}
\delta \int_0^T \left( \int_\Omega \frac{1}{2} \rho u \cdot u - e(\rho, \mathsf{s}) - \frac{1}{2\mu_0} B \cdot B \text{d} x \right) \text{d} t = 0,
\end{equation}
with constrained variations
\begin{subequations}
\label{eq:constrained_variations_MHD}
\begin{align}
\delta u & =\partial_t v + (u \cdot \nabla) v - (v \cdot \nabla) u, \\
\delta \rho & = - \nabla \cdot (\rho v), \\
\delta B & = - \nabla \times (B \times u) - (\nabla \cdot B) u, \\
\delta \mathsf{s} & = - \nabla \cdot (\mathsf{s} v),
\end{align}
\end{subequations}
for an arbitrary time-dependent vector field $v$. These constrained variations are best understood through an Euler--Poincar\'e reduction process \cite{Holm1998}, which transforms the variational principle on the Lie group of diffeomorphims into one on its Lie algebra of vector fields. In general, variational principles provide a systematic framework for describing complex systems (including but not limited to fluid ones) that involve multiple potentials and intricate couplings. They also automatically relate conservation laws to symmetries through Noether's theorem. Finally, variational formulations are useful to derive structure-preserving numerical schemes \cite{Marsden1994}. These ideas have also now become a popular guiding tool for the discretization of fluid systems \cite{GaGB2021, Gawlik2024, Gawlik2025, Carlier2025, Vazquez2020_GEEC}. Finite-Element formulations are particularly suitable in this context, since variational principles naturally lead to a weak form of the governing equations.

\paragraph{Geometric formulation.}
Another important aspect is the geometric nature of the different variables. In MHD, the variables $\rho$, $B$ and $\mathsf{s}$ used above are only coordinate representations of the mass density, magnetic induction, and entropy density. A more intrinsic description is given in the language of differential forms
\begin{subequations}
\begin{align}
\varrho & = \rho \,\text{d} x^1 \wedge \text{d} x^2 \wedge \text{d} x^3, \\
\beta & = B^3 \text{d} x^1 \wedge \text{d} x^2 + B^1 \text{d} x^2 \wedge \text{d} x^3 + B^2 \text{d} x^3 \wedge \text{d} x^1, \\
s & = \mathsf{s} \,\text{d} x^1 \wedge \text{d} x^2 \wedge \text{d} x^3.
\end{align}
\end{subequations}
Note that this correspondence depends on a metric, here given by the usual inner product on $\mathbb{R}^3$. With this new representation, the previous constrained variations \eqref{eq:constrained_variations_MHD} can now be simply expressed with Lie derivatives
\begin{subequations}
\begin{align}
\delta u & = \partial_t v - \pounds_v u, \\
\delta \varrho & = - \pounds_v \varrho, \\
\delta \beta & = - \pounds_v \beta, \\
\delta s & = - \pounds_v s.
\end{align}
\end{subequations}
It will be shown in this work that this geometric representation also allows to treat in a complete analogous way resistivity and heat transfer, as well as their respective associated boundary conditions. In both cases, the dissipation term can be written with the exterior derivative of the dual variable (e.g the temperature $0$-form $T$ or the $1$-form representation of the magnetic field $H = \frac{1}{\mu_0} B$), while boundary conditions are naturally expressed with the inclusion map $i:\partial \Omega \rightarrow \Omega$ and its pullback action on forms. Beyond its simplicity and elegance, the geometric formulation offers several advantages. It allows for an easy extension to non-euclidean geometries and to variables of different natures (e.g. differential forms, tensor fields or tensor densities). It also avoids using a metric where it is not needed \cite{Kanzo2007, Alonso2026}, leading to a cleaner and more transparent formulation. Exterior calculus can also be replicated at the discrete level \cite{Desbrun2005, Arnold2010}. Finally, geometry streamlines the studies of symmetry and its consequences on the equations.

\paragraph{Extension to non-equilibrium thermodynamics, closures, symmetries, and Curie's principle.}
The variational principle \eqref{eq:vp_MHD}-\eqref{eq:constrained_variations_MHD} accounts only for ideal flows; viscosity, resistivity, and heat transfer cannot be incorporated directly. In general, Hamilton's principle describes only reversible processes and is therefore incompatible with dissipative systems. This limitation is too restrictive, since most applications, such as those arising in Magnetic Confinement Fusion \cite{Doyle2007, Chen2016}, involve significant and intricate sources of dissipation. A new variational framework addressing this issue was introduced in \cite{GayBalmaz2017a, GayBalmaz2017b, Gay-BalmazYoshimura2019Noneq}. Dissipation is incorporated through the addition of both a phenomenological constraint and a variational constraint encoding all dissipative mechanisms. The resulting augmented principle shares similarities with the Lagrange--d'Alembert principle. The present article builds upon this framework by reformulating it in the language of differential forms. The entropy production equation, written as a volume form equation, displays the usual ``force-flux'' structure \cite{Groot1962}. Two types of fluxes will be of primary interest: the so-called discrete fluxes, typically relevant in chemical reactions, and continuous fluxes, which can model heat conduction, particle diffusion, or magnetic resistivity. Thermodynamically-consistent flux closures are then written and Onsager's principle is adapted to the present formalism. How these closures behave with respect to the symmetries of the system will be discussed and illustrated in the case of isotropy, while Curie’s principle is revisited within this geometric setting through the lens of representation theory.

\paragraph{Outline of the article.} In \S\ref{sec:notations} and \S\ref{sec:reversible} is summarized the classical geometric setting of variational principles for ideal fluid systems written with differential forms. The presentation of the new geometric variational principle for dissipative fluid systems is then divided into two parts. First, in \S\ref{sec:irreversible}, we provide a general metric-free formulation (both in its Lagrangian and reduced Eulerian formulation) for a fluid system evolving on an arbitrary oriented manifold, together with an advected differential form. Each dissipation mechanism is expressed in terms of a thermodynamic affinity and a thermodynamic flux. Equations will first be given in weak forms for a general dissipation process, and then in strong form for two specific closures on the thermodynamic affinities. Second, in \S\ref{sec:dissipation}, closures for thermodynamic fluxes are given. Onsager-like coupling between different dissipation mechanism are included. It is also shown how symmetry can simplify the different expressions through Curie's principle. Finally, \S\ref{sec:case_study} illustrates the methodology on a multi-species extension of equations \eqref{eq:MHD}, involving several interacting sources of dissipation.

\section{Physical description and notations}
\label{sec:notations}

\paragraph{Diffeomorphism group and vector fields.} Let $\Omega$ be an orientable manifold of dimension $n$, with a (potentially empty) smooth boundary. The smooth motion of a fluid within $\Omega$ can be viewed as a curve in $G = \operatorname{Diff}(\Omega)$, the group of diffeomorphisms of $\Omega$. The diffeomorphim $\varphi(t)$, at time $t$, keeps track of where the infinitesimal fluid elements are. If an infinitesimal fluid element is initially at $X \in \Omega$, it is now at $x = \varphi(t)(X) \in \Omega$. The group $G$ is formally interpreted as an infinite-dimensional Lie group \cite{Arnold1966, Holm1998}, with function composition as its group operation. The Lie algebra of $G$ is naturally identified with $\mathfrak{g} = \mathfrak{X}(\Omega)$, the space of smooth vector fields on $\Omega$ tangent to the boundary. The dual of $\mathfrak{g}$ is identified with the space of 1-form-valued densities, that is sections of the bundle $T^* \Omega \otimes \Lambda^n T^* \Omega$. The dual pairing is given by, for all $(u, m) \in \mathfrak{g} \times \mathfrak{g}^*$,
\begin{equation}\label{pairing_1}
\langle m, u \rangle = \int_\Omega m \cdot u.
\end{equation}
where $\cdot$ is the contraction of $u$ with the first (covector) component of $m$.

\paragraph{Differential forms.} The fluid system may further involve additional advected quantities, such as the mass density, the entropy density, or the magnetic field. These variables are assumed to take values in a vector space 
$V$ on which the pull-back by diffeomorphisms is well defined; this action is denoted by $A\cdot \varphi= \varphi^*A$, for $A\in V$, $\varphi\in G$. In this article, we focus on the case where the advected quantities are differential $k$-forms, that is, 
$V=\Lambda^k(\Omega)$ for some integer $k \in \mathbb{N}$. The pull-back defines a right linear action (or representation) of the group $G$ on $V$ as
\begin{equation}
A \cdot (\phi \circ \varphi) = (\phi \circ \varphi)^* A = \varphi^*(\phi^* A) = (A \cdot \phi) \cdot \varphi,
\end{equation}
for all $\phi, \varphi \in G, A \in V$. This action then induces a Lie algebra representation on $V$, which coincides with the Lie derivative with respect to a vector field
\begin{equation}
A \cdot u = \pounds_u A = \left. \frac{{\rm d}}{{\rm d} s}\right|_{s = 0} \varphi(s)^* A,
\end{equation}
where $\varphi(s)\in G$ is such that $\varphi(0) = \operatorname{id}$ and $\dot{\varphi}(0) = u \in \mathfrak{g}$. A pull-back and a Lie derivative can be defined on a space $V^\star$ in nondegenerate duality with $V$ thanks to the dual bracket $\langle \cdot , \cdot \rangle$
\begin{subequations}
\begin{align}
\label{eq:dual_pullback}
\langle \varphi^* B, A \rangle & = \langle B, (\varphi^{-1})^* A \rangle \\
\label{eq:dual_lie_derivative}
\langle \pounds_u B, A \rangle & = - \langle B, \pounds_u A \rangle 
\end{align}
\end{subequations}
for all $\varphi \in G, u \in \mathfrak{g}, A \in V, B \in V^*$. Finally, for any time-varying element $A$ of $V$ or $V^\star$, the following notation is introduced
\begin{equation}
\mathcal{D}_t A = \dot{A} + \pounds_u A.
\end{equation}
In the following, the dual space $V^*$ will be conveniently chosen as the space of $(n-k)$-forms with, for all $A \in \Lambda^k(\Omega), B \in \Lambda^{n-k}(\Omega)$, the nondegenerate duality pairing
\begin{equation}
\langle B, A \rangle = \int_\Omega B \wedge A,
\end{equation}
where we assume that an orientation has been fixed on the orientable manifold $\Omega$. It is readily checked that this identification is consistent with \eqref{eq:dual_pullback}-\eqref{eq:dual_lie_derivative} (i.e. the pullback and Lie derivative of $B$ is the same if it is regarded as $(n-k)$-form or as a dual element of the space of $k$-forms). 

\begin{remark}
\label{rem:variables}
The advected quantities must often satisfy additional conditions that must be preserved by the pull-back operation. For differential forms, such conditions include (1) the closedness condition ${\rm d} A = 0$ and (2) the boundary condition $i_{\partial\Omega \rightarrow \Omega}^* A=0$, where $i_{\partial\Omega \rightarrow \Omega}: \partial \Omega \rightarrow \Omega$ is the natural inclusion ($A$ is then said to be normal to the boundary). If $A$ is taken to be the magnetic induction field $(n-1)$-form and $\vec{B}$ its coordinate representation, they respectively give $\nabla \cdot \vec{B} = 0$ and $\vec{B} \cdot n = 0$.
\end{remark}

\begin{remark}
Although this is out of the scope of this article, an arbitrary space $V$ on which $G$ acts by left (or right) representation could also be considered as a generalization.
\end{remark}

\section{Reversible case}
\label{sec:reversible}

Before turning to the irreversible case and the treatment of entropic processes, we first review the reversible setting. The approach follows the Lie group variational formulation of fluid dynamics presented in \cite{Holm1998}, including its Euler–Poincaré reduction. The variational principle is, however, slightly adapted to incorporate the dual variable $B$, which plays an important role in the non-equilibrium theory; see \cite{ElGBWu2024} for its role in open fluid systems.

\subsection{Lagrangian formulation}


The variational principle is formulated in terms of the  Lagrangian of the system, defined as a map
\begin{equation}\label{fluid_Lagrangian}
L: TG \times V \rightarrow \mathbb{R}.
\end{equation}
It typically represents the difference between the kinetic and potential energies of the system. For most fluid models, the Lagrangian in \eqref{fluid_Lagrangian} is induced by a Lagrangian density $\mathscr{L}$ taking values in $\Lambda^n(\Omega)$, and can be written as
\begin{equation}\label{Lagr_dens}
L(\varphi,\dot{\varphi}, A)= \int_\Omega \mathscr{L}(\varphi(X), \dot{\varphi}(X), \nabla\varphi(X), A(X)).
\end{equation}
Here, $\nabla\varphi$ denotes the deformation gradient of $\varphi$, which is defined independently of any choice of Riemannian metric. From a differential-geometric perspective, one may instead view this quantity intrinsically as the tangent map $T_X\varphi: T_X\Omega\rightarrow T_{\varphi(X)}\Omega$ which encodes both the value $\varphi(X)$ and the linear map $\nabla \varphi(X)$.

\begin{varia}
Given a Lagrangian $L: TG \times V \rightarrow \mathbb{R}$, the variational problem consists in finding $\varphi: [0, T] \rightarrow G$, $A: [0, T] \rightarrow V$ and $B: [0, T] \rightarrow V^*$ such that
\begin{equation}
\label{eq:var1}
\delta \int_0^T L(\varphi(t), \dot{\varphi}(t), A(t)) + \big\langle \dot{B}(t), A(t) \big\rangle {\rm d} t = 0,
\end{equation}
for arbitrary variations $\delta \varphi$, $\delta A$, $\delta B$, with $\delta \varphi$ and $\delta B$ vanishing at endpoints $t = 0$ and $t = T$.
\end{varia}

\begin{remark}[Formulation with $n$-forms]
Using the Lagrangian density defined in \eqref{Lagr_dens}, the critical condition in \eqref{eq:var1} can be written (omitting the variables $X$ and $t$) as
\begin{equation}
\label{eq:var1_Omega2}
\delta \int_0^T \int_\Omega \mathscr{L}( \varphi, \dot{\varphi}, \nabla\varphi, A)  + \dot B \wedge A \, {\rm d} t = 0.
\end{equation}
The action functional is explicitly obtained by integrating an $n$-form over the fluid domain $\Omega$.
\end{remark}

\begin{remark}[Form of the constraint term]\label{rem:weak_dotA}
The dual variable $B$ acts as a Lagrange multiplier to enforce weakly the condition $\dot{A} = 0$. Although the stationarity of $A$ could have also been enforced strongly, it is here done weakly in order to lay the foundations for the irreversible case. The corollary condition $\dot{B} = - \frac{\delta L}{\delta A}$ will then be crucial for expressing the dissipation in the extended variational principle.
\end{remark}

\begin{remark}[Several advected variables]
This variational principle (as well as those presented in the following section) can be extended in a straightforward manner to include an arbitrary number of auxiliary variables, each belonging to its own space. For simplicity, we first consider a single such variable.
\end{remark}

\subsection{Reduction and Eulerian formulation}

\paragraph{Reduction of the Lagrangian.}
We assume that the Lagrangian satisfies the relabelling symmetry
\begin{equation}
\label{eq:symmetry_L}
L(\varphi \circ \phi , \dot{\varphi} \circ\phi, \phi^* A ) = L(\varphi, \dot{\varphi}, A), \quad \text{for all $\phi \in G$},
\end{equation}
that is, $L$ is invariant under the right action of the group $G$.
Under this invariance assumption, the variational principle \eqref{eq:var1} can be reformulated entirely in Eulerian variables. Choosing $\phi = \varphi^{-1}$, and introducing the Eulerian velocity $u = \dot{\varphi} \circ \varphi^{-1}$ together with the advected quantity $a = \varphi_* A = (\varphi^{-1})^* A$, we define the reduced Lagrangian
\begin{equation}
\ell (u, a) = L(\operatorname{id}, u, a).
\end{equation}
The invariance in \eqref{eq:symmetry_L} typically comes from the so-called material covariance of the Lagrangian density $\mathscr{L}$ \cite{GBMaRa2012,GB2024} which reads as follows
\begin{equation}
\mathscr{L}(\varphi \circ \phi, \varphi \circ \phi, \nabla (\varphi \circ \phi), \phi^* A)= \phi^*\left[\mathscr{L}(\varphi, \dot\varphi, \nabla\varphi, A) \right], \quad \text{for all $\phi \in G$}.
\end{equation}
From this, one obtains the spatial Lagrangian density $\mathfrak{l}$, taking values in $\Lambda^n(\Omega)$, such that 
\begin{equation}
\mathscr{L}(\varphi, \dot\varphi, \nabla \varphi, A)= \varphi^*[\mathfrak{l}(u, a)], \quad \text{for $u=\dot\varphi\circ\varphi^{-1}$ and $a=\varphi_*A$.}
\end{equation}
Eventually, the Eulerian version of \eqref{Lagr_dens} is given as
\begin{equation}\label{Lagr_dens_Eul}
\ell(u,A)=\int_\Omega\mathfrak{l}(u(x), A(x)).
\end{equation}

\paragraph{Reduction of the constraint.}
The constraint appearing in the second term of \eqref{eq:var1} can likewise be rewritten in Eulerian form in terms of the advected variables of $a=\varphi_*A$ and $b = \varphi_* B$. To do so, first note that
\begin{subequations}
\label{eq:lemma1}
\begin{align}
\left. \frac{{\rm d}}{{\rm d} t} \right|_{t = t_0} \left( \varphi^{-1}(t)^{*} B(t) \right) 
& = \left. \frac{{\rm d}}{{\rm d} t} \right|_{t = t_0} \left( (\varphi^{-1}(t) \circ \varphi(t_0) \circ \varphi^{-1}(t_0))^{*} B(t) \right) \\
& = \varphi^{-1} (t)^* \left. \frac{{\rm d}}{{\rm d} t} \right|_{t = t_0} B(t) - \pounds_{\dot{\varphi}(t_0) \circ \varphi^{-1}(t_0)} \left( \varphi^{-1}(t_0)^* B(t_0) \right),
\end{align}
\end{subequations}
so that
\begin{equation}
\left \langle \dot{B}, A \right \rangle 
= \left \langle (\varphi^{-1})^* \dot{B}, (\varphi^{-1})^* A \right \rangle
= \left \langle \dot{b} + \pounds_u b, a \right \rangle
= \left \langle \mathfrak{D}_t b, a \right \rangle.
\end{equation}
Finally, since the Eulerian velocity $u$ is expressed in terms of $\varphi$, its variation $\delta u$ is induced by variations $\delta\varphi$. More precisely, $\delta u$ depends on the Eulerian variation $v = \delta \varphi \circ \varphi^{-1}$. For completeness, we provide the derivation of the constrained variation $\delta u$. Using local coordinates
\begin{subequations}
\begin{align}
\frac{\partial}{\partial s} \left( \left( \frac{\partial}{\partial t} \varphi(t, s) \right) \circ \varphi^{-1}(t, s) \right) & = \left( \frac{\partial}{\partial s} \frac{\partial}{\partial t} \varphi \right) \circ \varphi^{-1} + \frac{\partial}{\partial X^i} \left( \frac{\partial}{\partial t} \varphi \right) \circ \varphi^{-1}  \frac{\partial}{\partial s} (\varphi^{-1})^i \\
& = \left( \frac{\partial}{\partial s} \frac{\partial}{\partial t} \varphi \right) \circ \varphi^{-1} - \frac{\partial}{\partial X^i} \left( \frac{\partial}{\partial t} \varphi \right) \circ \varphi^{-1} \frac{\partial}{\partial x^j} (\varphi^{-1})^i \left( \frac{\partial}{\partial s} \varphi_j \right) \circ \varphi^{-1} \\
& = \left( \frac{\partial}{\partial s} \frac{\partial}{\partial t} \varphi \right) \circ \varphi^{-1} - \frac{\partial}{\partial x_j} \left( \left( \frac{\partial}{\partial t} \varphi \right) \circ \varphi^{-1} \right) \left( \frac{\partial}{\partial s} \varphi^j \right) \circ \varphi^{-1}
\end{align}
\end{subequations}
which gives both (permuting the role of $t$ and $s$)
\begin{subequations}
\begin{align}
\delta u & = \left( \frac{\partial}{\partial s} \frac{\partial}{\partial t} \varphi \right) \circ \varphi^{-1} - v^j \frac{\partial}{\partial x^j} u \\
\dot{v} & = \left( \frac{\partial}{\partial s} \frac{\partial}{\partial t} \varphi \right) \circ \varphi^{-1} - u^j \frac{\partial}{\partial x^j} v
\end{align}
\end{subequations}
so that eventually
\begin{equation}
\delta u = \dot{v} + u^j \frac{\partial}{\partial x^j} v - v^j \frac{\partial}{\partial x^j} u = \dot{v} + [u,v].
\end{equation}

\begin{varia}
Given a reduced Lagrangian $\ell: \mathfrak{g} \times V \rightarrow \mathbb{R}$, the reduced variational problem consists in finding curves $u: [0, T] \rightarrow \mathfrak{g}$, $a: [0, T] \rightarrow V$ and $b: [0, T] \rightarrow V^*$ such that
\begin{equation}
\label{eq:var2}
\delta \int_0^T \ell(u(t), a(t)) + \left\langle \mathcal{D}_t b(t), a(t) \right\rangle {\rm d} t = 0,
\end{equation}
where $\delta u = \partial_t v + [u,v]$, with $v :[0,T]\rightarrow \mathfrak{g}$, and where $v$ and $\delta b$ vanish at endpoints $t=0$ and $t = T$.
\end{varia}

\begin{remark}[Formulation with $n$-forms]
Using the Lagrangian density defined in \eqref{Lagr_dens_Eul}, the critical condition in \eqref{eq:var2} can be written (omitting the variables $x$ and $t$) as
\begin{equation}
\label{eq:var1_Omega3}
\delta \int_0^T \int_\Omega \mathfrak{l}(u,a)  + \mathcal{D}_t b  \wedge a \, {\rm d} t = 0.
\end{equation}
Again, the action functional is obtained by integrating an $n$-form over the fluid domain $\Omega$.
\end{remark}

\paragraph{Resulting equations.} The principle \eqref{eq:var2} yields
\begin{equation}
\int_0^T 
  \left \langle \frac{\delta \ell}{\delta u}, \delta u \right \rangle 
+ \left \langle \frac{\delta \ell}{\delta a}, \delta a \right \rangle - \left \langle \delta b, \mathcal{D}_t a \right \rangle 
+ \left \langle \pounds_{\delta u} b, a\right \rangle
+ \left \langle \mathcal{D}_t b, \delta a\right \rangle {\rm d} t = 0,
\end{equation}
where $\frac{\delta \ell}{\delta u}$ is a 1-form density over $\Omega$, seen as the dual of vector fields. Free variation with respect to $a$ and $b$ immediately gives
\begin{equation}
\label{eq:dtb_dta}
\mathcal{D}_t b = - \frac{\delta \ell}{\delta a}, \quad \quad \quad
\mathcal{D}_t a = 0.
\end{equation}
Using the constrained variations on $u$, there remains
\begin{equation}
\int_0^T 
\left \langle \frac{\delta \ell}{\delta u}, \dot{v} + [u, v] \right \rangle
+ \left \langle \pounds_{\dot{v} + [u,v]} b, a \right \rangle {\rm d} t = 0.
\end{equation}
Because $\pounds_{[u,v]} = \pounds_u \pounds_v -  \pounds_v \pounds_u$, the second term can be written
\begin{subequations}
\label{eq:simplification}
\begin{align}
\left \langle \pounds_{\dot{v} + [u,v]} b, a \right \rangle 
& = 
- \langle \pounds_v \dot{b}, a \rangle
- \left \langle \pounds_v b, \dot{a} \right \rangle
+ \left \langle \pounds_u \pounds_v b, a \rangle 
- \langle \pounds_v \pounds_u b, a \right \rangle \\
& = 
+ \langle \dot{b}, \pounds_v a \rangle
- \left \langle \pounds_v b, \dot{a} \right \rangle
- \left \langle \pounds_v b, \pounds_u a \rangle 
+ \langle \pounds_u b, \pounds_v a \right \rangle \\
& = 
\langle \mathcal{D}_t b, \pounds_v a \rangle
- \langle \pounds_v b, \mathcal{D}_t a \rangle.
\end{align}
Eventually, using \eqref{eq:dtb_dta}, free variations on $v$ yields the strong form of the momentum equation
\begin{equation}
\frac{\partial}{\partial t} \frac{\delta \ell}{\delta u} + \pounds_u \frac{\delta \ell}{\delta u} - \frac{\delta \ell}{\delta a} \diamond a = 0,
\end{equation}
\end{subequations}
where $\frac{\delta \ell}{\delta a} \diamond a \in \mathfrak{g}^*$ is the 1-form density defined by $\langle \frac{\delta \ell}{\delta a} \diamond a, v \rangle_\mathfrak{g} = - \langle \frac{\delta \ell}{\delta a}, \pounds_v a \rangle_V$ for all $v \in \mathfrak{g}$. It is possible to derive an explicit form using the fact that
\begin{equation}
\label{eq:IPP}
- \int_\Omega b \wedge \pounds _v a = \int_\Omega (-1)^{n-k} {\rm d}b \wedge {\rm i}_v a - b\wedge {\rm i}_v{\rm d}a - (-1)^{n-k} \int_{\partial \Omega} i^*_{\partial \Omega \rightarrow \Omega} (b \wedge {\rm i}_v a),
\end{equation}
so that, by non-degeneracy of the coupling,
\begin{equation}
\label{eq:general_diamond}
b \diamond a = (-1)^{n-k} {\rm d}b \wedge {\rm i}_{\,\_\,} a - b\wedge {\rm i}_{\,\_\,} {\rm d}a= {\rm i}_{\,\_\,} {\rm d} b \wedge a - b\wedge {\rm i}_{\,\_\,} {\rm d}a.
\end{equation}
Note that for \eqref{eq:general_diamond} to be true, the boundary term in \eqref{eq:IPP} must vanish. In general such boundary term disappears, either because a zero velocity is imposed at the boundary (the variational principle is restricted to the subgroup of pointwise boundary-preserving diffeomorphisms), or because the advected quantity satisfies the boundary condition $i^*_{\partial \Omega \rightarrow \Omega} a = 0$. This is related to Remark \ref{rem:variables}.

\begin{remark}
In dimension $n=3$ with the usual scalar product, the explicit expression of $\pounds_u a$, $i^*_{\partial \Omega \rightarrow \Omega} a = 0$, and $\frac{\delta \ell}{\delta a} \diamond a$ can be given for forms of all degrees. In local coordinates, with 
\begin{subequations}
\begin{align}
f & \in \Lambda^0(\Omega), \\
\alpha = A_1 \text{d} x^1 + A_2 \text{d} x^2 + A_3 \text{d} x^3 & \in \Lambda^1(\Omega), \\
\beta = B_3 \text{d} x^1 \wedge \text{d} x^2 + B_1 \text{d} x^2 \wedge \text{d} x^3 + B_2 \text{d} x^3 \wedge \text{d} x^1 & \in \Lambda^2(\Omega), \\
\varrho = \rho \text{d} x^1 \wedge \text{d} x^2 \wedge \text{d} x^3 & \in \Lambda^3(\Omega),
\end{align}
\end{subequations}
Lie derivatives are then expressed as
\begin{subequations}
\begin{align}
\pounds_u f & = u \cdot \nabla f, \\
\pounds_u \alpha & = \left[(\nabla \times A) \times u + \nabla (A \cdot u) \right] \cdot (\text{d}x^1, \text{d}x^2, \text{d}x^3), \\
\pounds_u \beta & = \left[\nabla \times (u \times B) + u \nabla \cdot B\right] \cdot (\text{d} x^2 \wedge \text{d} x^3, \text{d} x^3 \wedge \text{d} x^1, \text{d} x^1 \wedge \text{d} x^2), \\
\pounds_u \varrho & = \nabla \cdot (\rho u) \text{d} x^1 \wedge \text{d} x^2 \wedge \text{d} x^3,
\end{align}
\end{subequations}
while the boundary conditions can be expressed with the outward normal vector field $n$ as
\begin{subequations}
\begin{align}
i^*_{\partial \Omega \rightarrow \Omega} f & = 0 & \Longleftrightarrow & & f = 0, \\ 
i^*_{\partial \Omega \rightarrow \Omega} \alpha & = 0 & \Longleftrightarrow & & A \times n = 0, \\ 
i^*_{\partial \Omega \rightarrow \Omega} \beta & = 0 & \Longleftrightarrow & & B \cdot n = 0, \\ 
i^*_{\partial \Omega \rightarrow \Omega} \varrho & = 0 & \Longleftrightarrow & & \varnothing.
\end{align}
\end{subequations}
Finally, under cancellation of the boundary term in \eqref{eq:IPP},
\begin{subequations}
\begin{align}
\varrho \diamond f & = - f \diamond \varrho = - (\rho \nabla f)^\flat \text{d} x^1 \wedge \text{d} x^2 \wedge \text{d} x^3, \\
\alpha \diamond \beta & = - \beta \diamond \alpha = \left[ (\nabla \times A) \times B + (\nabla \cdot B) A \right]^\flat \text{d} x^1 \wedge \text{d} x^2 \wedge \text{d} x^3.
\end{align}
\end{subequations}    
\end{remark}

\paragraph{MHD equations.} The previous principle may be applied to a subdomain $\Omega \subset \mathbb{R}^3$, with three variables: the mass density $a_1 = \varrho \in \Lambda^3(\Omega)$, the entropy density $a_2 = s \in \Lambda^3(\Omega)$, and the magnetic field $a_3 = \beta \in \Lambda^2(\Omega)$, with $i^*_{\partial \Omega \rightarrow \Omega} a_i = 0$ for all $1 \leq i \leq n$ (note that this gives no boundary condition for $\varrho$ and $s$). It eventually produces the equations
\begin{subequations}
\begin{align}
\frac{\partial}{\partial t} \frac{\delta \ell}{\delta u} + \pounds_u \frac{\delta \ell}{\delta u} & = \frac{\delta \ell}{\delta \varrho} \diamond \varrho + \frac{\delta \ell}{\delta s} \diamond s + \frac{\delta \ell}{\delta \beta} \diamond \beta, \\
\mathcal{D}_t \varrho & = 0, \\
\mathcal{D}_t s & = 0, \\
\mathcal{D}_t \beta & = 0.
\end{align}
\end{subequations}
Given the usual metric on $\Omega$, and writing $\rho = \star \varrho$, $\mathsf{s} = \star s$, $B = (\star \beta)^\sharp$, $e(\rho, \mathsf{s}) = \star \epsilon(\varrho, s)$, and the Lagrangian
\begin{equation}
\ell(u, \varrho, s, \beta) = \int_\Omega \frac{1}{2} (u \cdot u) \varrho - \epsilon(\rho, s) - \frac{1}{2 \mu_0} \beta \wedge \star \beta = \int_\Omega \frac{1}{2} \rho (u \cdot u) - e(\rho, \mathsf{s}) - \frac{1}{2\mu_0} B \cdot B {\rm d} x,
\end{equation}
one recovers the ideal MHD equations (i.e. \eqref{eq:MHD} with $\sigma = 0$, $\kappa = \nu = 0$, and with the boundary condition $B \cdot n = 0$).

\section{Irreversible case}
\label{sec:irreversible}

In this section, the previous variational principle is extended to account for non-equilibrium thermodynamics. Dissipation processes are described with thermodynamic forces and thermodynamic fluxes \cite{Onsager1931a, Onsager1931b}. Consistently with the rest of the article, both are expressed as differential forms.

\subsection{Lagrangian formulation}

\paragraph{Entropy.} Dissipation is accounted for through the entropy density $S$. Such entropy density is represented by a volume form, that is, $S \in \Lambda^n(\Omega)$. The Lagrangian now also depends on $S$, namely
\begin{equation}
L(\varphi, \dot \varphi, A, S)\quad \text{where}\quad  L: TG \times \Lambda^k(\Omega) \times \Lambda^n(\Omega) \rightarrow \mathbb{R}.
\end{equation}
The temperature is defined as the opposite of the dual variable $T = -\delta L/\delta S \in \Lambda^0(\Omega)$. In accordance with the second law of thermodynamics, the entropy density is decomposed as
\begin{equation}\label{decomposed_entropy}
S = \Sigma + (S - \Sigma),
\end{equation}
where $\Sigma \in \Lambda^n(\Omega)$ represents the produced entropy and $S - \Sigma$ corresponds to the exchanged entropy. 

\paragraph{Dissipation in terms of differential forms.} The rate of total entropy production $\dot \Sigma$ is assumed to be the sum of contributions of the form
\begin{equation}
\frac{1}{T} X \wedge J \in \Lambda^n(\Omega),
\end{equation}
where $X \in \Lambda^{\ell}(\Omega)$ denotes a thermodynamic affinity (or force) and $J \in \Lambda^{n-\ell}(\Omega)$ is the associated thermodynamic flux. This expression represents the familiar “force–flux” structure of entropy production in nonequilibrium thermodynamics \cite{Groot1962}, formulated here intrinsically in the language of differential forms. The thermodynamic affinities are also assumed to be linear functions of the dual variable $X(\frac{\delta L}{\delta A})$ where $X:\Lambda^{n-k}(\Omega) \rightarrow \Lambda^{\ell}(\Omega)$ commutes with the action of $G$ by pullback, i.e.
\begin{equation}\label{X_equivariance}
\phi^* (X(F)) = X( \phi^*F), \quad\text{for all $\phi\in G$}.
\end{equation}
Note however that $J$ does not need to be linear in the dual variables. In the variational principle, the entropy production is given in weak form, thanks to the trilinear form $d_{A}:\Lambda^0(\Omega) \times \Lambda^{n-\ell}(\Omega) \times \Lambda^{n-k}(\Omega)$
\begin{equation}
d_{A}\left(W, J, F\right) = \int_\Omega W X(F) \wedge J.
\end{equation}
From \eqref{X_equivariance} and the property of the pull-back together with the change-of-variables formula, the dissipation function $d_A$ satisfies the invariance property
\begin{equation}\label{d_A_inv}
d_{A}\left(\phi^*W, \phi^*J, \phi^*F\right)= d_{A}\left(W, J, F\right), \quad\text{for all $\phi\in G$}.
\end{equation}
In practice, two categories of dissipation will be considered: a discrete and a continuous version
\begin{subequations}
\label{eq:two_fluxes}
\begin{align}
\label{eq:discrete_flux}
d^d(W, J, F) & = \int_{\Omega} W F \wedge J, \\
\label{eq:continuous_flux}
d^c(W, J, F) & = \int_{\Omega} W {\rm d}F \wedge J.
\end{align}
\end{subequations}
corresponding to the two choices $X(F)=F$ ($\ell=n-k$) and $X(F)= {\rm d}F$ ($\ell=n-k+1$), for which \eqref{X_equivariance} trivially holds. As we shall see later, they will respectively yield the following type of equations for the $k$-form $A$:
\begin{subequations}
\begin{align}
\dot{A} & = J, \\
\dot{A} & = \pm {\rm d} J.
\end{align}
\end{subequations}
The first case typically models chemical reactions but could also be used for heat transfer in two-entropy systems, or drag in two-velocity systems (once extended to vector-valued variables); the second case encompasses heat dissipation, mass dissipation and resistivity for magnetohydrodynamics (MHD) applications. These two choices are studied in details in \S\ref{sec:dissipation}, together with a comment on how viscosity can be added in a formally equivalent manner. In the meantime, the formulation of the variational principle will only be given in weak form with an arbitrary dissipation functions $d_A$. This allows to focus on the common algebraic features of all dissipation processes, automatically include potential boundary conditions, as well as being consistent with structure-preserving finite element methods \cite{Gawlik2024, Gawlik2025}.

Following \cite{GayBalmaz2017b}, dissipation is incorporated into the variational principle by means of both a phenomenological constraint and a variational constraint. The former expresses the rate of entropy production $\dot \Sigma$ as the sum of contributions arising from irreversible processes, while the latter provides the corresponding constraint on the variations $\delta \Sigma$. The resulting variational principle is an example of a generalized Lagrange-d'Alembert principle, in which entropy production enters through a nonholonomic constraint.

\begin{varia}
\label{vp:irreversible_lagrangian}
Given a Lagrangian $L: TG \times \Lambda^k(\Omega) \times \Lambda^n(\Omega) \rightarrow \mathbb{R}$, the variational problem consists in finding $\varphi: [0, T] \rightarrow G$, $S: [0, T] \rightarrow \Lambda^n (\Omega)$, $\Sigma: [0, T] \rightarrow \Lambda^n (\Omega)$, $\Gamma: [0, T] \rightarrow \Lambda^0 (\Omega)$, $A: [0, T] \rightarrow \Lambda^k(\Omega)$ and $B: [0, T] \rightarrow \Lambda^{n-k}(\Omega)$ such that
\begin{equation}\label{critical_cond3}
\delta \int_0^T L(\varphi(t), \dot{\varphi}(t), A(t), S(t)) + \big\langle \dot{B}(t), A(t) \big\rangle + \big\langle \dot{\Gamma}(t), S(t) - \Sigma(t) \big\rangle {\rm d} t = 0
\end{equation}
with the phenomenological and variational constraints
\begin{subequations}
\begin{align}
\label{eq:pheno_constraint_L}
\langle W, \dot{\Sigma} \rangle 
& = 
d_A \left(\frac{W}{\frac{\delta L}{\delta S}}, J_A, \dot{B} \right), & \forall \ W \in \Lambda^0(\Omega), \\
\label{eq:varia_constraint_L}
\left\langle W, \delta \Sigma \right\rangle 
& = 
d_A \left(\frac{W}{\frac{\delta L}{\delta S}}, J_A, \delta B \right), & \forall \ W \in \Lambda^0(\Omega),
\end{align}
\end{subequations}
and shuch that variations $\delta \Phi$ and $\delta B$ vanish at endpoints $t = 0$ and $t = T$.
\end{varia}

\begin{remark}[Formulation with $n$-forms]
Using the Lagrangian density defined in \eqref{Lagr_dens}, the critical condition in \eqref{critical_cond3}--\eqref{eq:pheno_constraint_L}--\eqref{eq:varia_constraint_L} can be written as
\begin{equation}
\label{eq:var1_Omega1}
\delta \int_0^T \int_\Omega \mathscr{L}( \varphi, \dot{\varphi}, \nabla\varphi, A)  + \dot B \wedge A + \dot \Gamma \wedge (S-\Sigma)\, {\rm d} t = 0
\end{equation}
with the phenomenological and variational constraints
\begin{subequations}
\begin{align}
\label{eq:pheno_constraint_L_n_form}
\int_\Omega W\wedge\dot{\Sigma} 
& = 
\int_\Omega\frac{W}{\frac{\delta L}{\delta S}}\wedge \dot{B}\wedge J_A,
& \text{resp.} \quad \quad & 
\int_\Omega\frac{W}{\frac{\delta L}{\delta S}}\wedge {\rm d}\dot{B}\wedge J_A, & \forall \; W \in \Lambda^0(\Omega), \\
\label{eq:varia_constraint_L_n_form}
\int_\Omega W \wedge\delta \Sigma
& = 
\int_\Omega\frac{W}{\frac{\delta L}{\delta S}}\wedge \delta B\wedge J_A,
& \text{resp.} \quad \quad & 
\int_\Omega\frac{W}{\frac{\delta L}{\delta S}}\wedge {\rm d}\delta{B}\wedge J_A, & \forall \; W \in \Lambda^0(\Omega),
\end{align}
\end{subequations}
where in \eqref{eq:pheno_constraint_L_n_form} and \eqref{eq:varia_constraint_L_n_form} are given the discrete (resp. continuous) version of the dissipation function. 
\end{remark}

\begin{remark}
In \eqref{eq:pheno_constraint_L}, $\dot B$ is where the dual variable $\frac{\delta L}{\delta A}$ should be. This is consistent with Remark \ref{rem:weak_dotA} and the fact that $\dot{B} = - \frac{\delta L}{\delta A}$. As usual with d'Alembert-type variational principles, \eqref{eq:varia_constraint_L} is obtained by replacing time derivatives with variations, hence the use of $\delta B$.
\end{remark}

\subsection{Reduction and Eulerian formulation}

To derive a reduced version of the variational principle \ref{vp:irreversible_lagrangian}, the symmetry condition \eqref{eq:symmetry_L} must be extended to include the entropy variable, namely,
\begin{equation}
\label{eq:symmetry_L2}
L(\varphi \circ \phi , \dot{\varphi} \circ \phi, \phi^* A , \phi^*  S ) = L(\varphi, \dot{\varphi}, A, S), \quad\text{for all $\phi\in G$}.
\end{equation}
This invariance yields a reduced Lagrangian depending on the Eulerian entropy density $s = \varphi_* S$,
\begin{equation}
\ell(u,a,s)= L({\rm id}, u, a,s).
\end{equation}
The Eulerian variables $w=\varphi_*W$, $\varsigma= \varphi_* \Sigma$ and $\gamma = \varphi_* \Gamma$ are also introduced. The left-hand side of \eqref{eq:pheno_constraint_L} then becomes
\begin{equation}
\big\langle W, \dot{\Sigma} \big\rangle = \big\langle \varphi_* W, \varphi_* \dot{\Sigma} \big\rangle = \big\langle w, \mathcal{D}_t \varsigma \big\rangle,
\end{equation}
Using the invariance property \eqref{d_A_inv}, the right-hand side of \eqref{eq:pheno_constraint_L} can also be rewritten in Eulerian form as
\begin{equation}
\label{eq:symmetry_d}
d_A \left( \frac{W}{\frac{\delta L}{\delta S}}, J_A, \dot{B} \right) 
= 
d_A\left( \varphi_* \frac{W}{\frac{\delta L}{\delta S}}, \varphi_* J_A, \varphi_* \dot{B} \right) 
= 
d_A\left(\frac{w}{\frac{\delta \ell}{\delta s}}, j_a, \mathcal{D}_t b \right),
\end{equation}
where $j_a=\varphi_* J_A$ is the Eulerian thermodynamic flux, and we used the identity $\varphi_* \frac{\delta L}{\delta S}= \frac{\delta \ell}{\delta s}$, which follows from the relation between $L$ and $\ell$, and from $s=\varphi_*S$. In the following, the notation $d_a$ will be used as a substitute for $d_A$. Although both represent the same function, this change of notation emphasizes the context in which they are used. As for \eqref{eq:varia_constraint_L}, its Eulerian counterpart is naturally found by introducing the operator
\begin{equation}
\mathcal{D}_\delta a = \delta a + \pounds_v a,
\end{equation}
where, as before, $v = \delta \varphi \circ \varphi^{-1}$. While $\mathcal{D}_t a$ represents the Eulerian version of $\dot{A}$, the operator $\mathcal{D}_\delta a$ corresponds to the Eulerian version of $\delta A$. The Eulerian formulation of the variational principle \ref{vp:irreversible_lagrangian} can now be given.

\begin{varia}
\label{vp:4}
Given a reduced Lagrangian $\ell: \mathfrak{g} \times \Lambda^k(\Omega) \times \Lambda^n(\Omega) \rightarrow \mathbb{R}$, the variational problem consists in finding $u: [0, T] \rightarrow \mathfrak{g}$, $s: [0, T] \rightarrow \Lambda^n (\Omega)$, $\varsigma: [0, T] \rightarrow \Lambda^n (\Omega)$, $\gamma: [0, T] \rightarrow \Lambda^0 (\Omega)$, $a: [0, T] \rightarrow \Lambda^k(\Omega)$ and $b: [0, T] \rightarrow \Lambda^{n-k}(\Omega)$ such that
\begin{equation}
\label{eq:var4}
\delta \int_0^T \ell(u(t), a(t), s(t)) + \left\langle \mathcal{D}_t b (t), a(t) \right\rangle + \left\langle \mathcal{D}_t \gamma (t), s(t) - \varsigma(t) \right\rangle {\rm d} t = 0
\end{equation}
with the phenomenological and variational constraints
\begin{subequations}
\begin{align}
\label{eq:eulerian_phenomemo_constraint}
\left\langle w, \mathcal{D}_t \varsigma \right\rangle 
& =
d_a \left( \frac{w}{\frac{\delta \ell}{\delta s}}, j_a, \mathcal{D}_t b \right), \quad \quad \forall w \in \Lambda^0(\Omega), \\
\label{eq:eulerian_variational_constraint}
\left\langle w, \mathcal{D}_\delta \varsigma \right\rangle
& = 
d_a \left( \frac{w}{\frac{\delta \ell}{\delta s}}, j_a, \mathcal{D}_\delta b \right), \quad \quad \forall w \in \Lambda^0(\Omega).
\end{align}
\end{subequations}
where $\delta u = \partial_t v + [u,v]$, with $v: [0,T] \rightarrow \mathfrak{g}$, and where $v$, $\delta b$ and $\delta \gamma$ all vanish at endpoints $t=0$ and $t=T$.
\end{varia}

\begin{remark}[Formulation with $n$-forms]
Using the Lagrangian density defined in \eqref{Lagr_dens_Eul}, the critical condition in \eqref{eq:var4}--\eqref{eq:eulerian_phenomemo_constraint}--\eqref{eq:eulerian_variational_constraint} can be written as
\begin{equation}
\label{eq:var1_Omega4}
\delta \int_0^T \int_\Omega \mathfrak{l}( u,a,s)  + \mathcal{D}_tb \wedge a + \mathcal{D}_t \gamma \wedge (s-\varsigma)\, {\rm d} t = 0
\end{equation}
with the phenomenological and variational constraints
\begin{subequations}
\begin{align}
\label{eq:pheno_constraint_L_n_form4}
\int_\Omega w\wedge \mathcal{D}_t\varsigma 
& = 
\int_\Omega\frac{w}{\frac{\partial \mathfrak{l}}{\partial s}}\wedge \mathcal{D}_t b\wedge j_a  ,\quad\text{resp.}\quad  \int_\Omega\frac{w}{\frac{\partial \mathfrak{l}}{\partial s}}\wedge {\rm d}\mathcal{D}_tb\wedge j_a, \quad \forall \; w \in \Lambda^0(\Omega), \\
\label{eq:varia_constraint_L_n_form4}
\int_\Omega w \wedge \mathcal{D}_\delta\varsigma
& = 
\int_\Omega\frac{w}{\frac{\partial \mathfrak{l}}{\partial s}}\wedge \mathcal{D}_\delta b\wedge j_a,\quad\text{resp.}\quad  \int_\Omega\frac{w}{\frac{\partial \mathfrak{l}}{\partial s}}\wedge {\rm d}\mathcal{D}_\delta b\wedge j_a, \quad \forall \; w \in \Lambda^0(\Omega),
\end{align}
\end{subequations}
where in \eqref{eq:pheno_constraint_L_n_form4} and \eqref{eq:varia_constraint_L_n_form4} are given the discrete (resp. continuous) version of the dissipation function. 
\end{remark}

\paragraph{Resulting equations in weak form.} Following the reduced reversible case, equation \eqref{eq:var4} yields
\begin{multline}
\int_0^T 
  \left \langle \frac{\delta \ell}{\delta u}, \delta u \right \rangle 
+ \left \langle \frac{\delta \ell}{\delta a}, \delta a \right \rangle + \left \langle \frac{\delta \ell}{\delta s}, \delta s \right \rangle - \left \langle \delta b, \mathcal{D}_t a \right \rangle 
+ \left \langle \pounds_{\delta u} b, a\right \rangle
+ \left \langle \mathcal{D}_t b, \delta a\right \rangle \\
- \left \langle \delta \gamma, \mathcal{D}_t (s - \varsigma) \right \rangle 
+ \left \langle \pounds_{\delta u} \gamma, s - \varsigma \right \rangle
+ \left \langle \mathcal{D}_t \gamma, \delta s - \delta \varsigma \right \rangle
{\rm d} t = 0.
\end{multline}
Free variation with respects to $a$ and $s$ immediately gives
\begin{equation}
\mathcal{D}_t b = - \frac{\delta \ell}{\delta a},
\quad \quad \text{and} \quad \quad
\mathcal{D}_t \gamma = - \frac{\delta \ell}{\delta s}.
\end{equation}
Using the variational constraint \eqref{eq:eulerian_variational_constraint}, the remaining terms read
\begin{multline}
\int_0^T 
\left \langle \frac{\delta \ell}{\delta u}, \delta u \right \rangle
+ \left \langle \pounds_{\delta u} b, a \right \rangle
- \left \langle \delta b, \mathcal{D}_t a \right \rangle
+ \left \langle \pounds_{\delta u} \gamma, s - \varsigma \right \rangle
\\
- \left \langle \delta \gamma, \mathcal{D}_t (s - \varsigma)
\right \rangle
+ \left \langle \mathcal{D}_t \gamma, \pounds_v \varsigma \right \rangle
\textcolor{blue}{+} d_a(1, j_a, \mathcal{D}_\delta b)
{\rm d} t = 0.
\end{multline}
Free variations with respect to $b$ and $\gamma$ then gives the conditions
\begin{subequations}
\begin{align}
\langle \delta b, \mathcal{D}_t a \rangle & = 
d_a\left( 1, j_a, \delta b \right), & \forall \; \delta b \in \Lambda^{n-k}(\Omega), \\
\langle w, \mathcal{D}_t (s - \varsigma) \rangle & = 0, & \forall\; w \in \Lambda^{0}(\Omega).
\end{align}
\end{subequations}
Using the constrained variations on $u$ and the simplification \eqref{eq:simplification} (as well as its counterpart replacing $a$ with $s - \varsigma$, and $b$ with $\gamma$), the weak equation for momentum eventually becomes
\begin{equation}
\label{eq:weak_form_momentum}
\left\langle \frac{\partial }{\partial t}\frac{\delta \ell}{\delta u}, v\right\rangle - \left\langle \frac{\delta \ell}{\delta u}, \pounds_uv\right\rangle + \left\langle \frac{\delta \ell}{\delta a}, \pounds_va \right\rangle+ \left\langle \frac{\delta \ell}{\delta s}, \pounds_vs \right\rangle=0, \quad \forall\; v\in\mathfrak{g}.
\end{equation}
The resulting equations in weak form thus consist of \eqref{eq:weak_form_momentum} together with
\begin{subequations}
\begin{align}
\left\langle w, \mathcal{D}_t s \right\rangle 
& =
d_a \left( \frac{w}{\frac{\delta \ell}{\delta s}}, j_a, \mathcal{D}_t b \right), 
& \forall \; w \in \Lambda^0(\Omega), \\
\langle \delta b, \mathcal{D}_t a \rangle 
& = 
d_a \left(1, j_a, \delta b \right), 
& \forall \;\delta b \in \Lambda^{n-k}(\Omega).
\end{align}
\end{subequations}

\paragraph{Inclusion of heat flux.}
In the variational principle \ref{vp:4}, the $n$-form $s-\varsigma \in \Lambda^{n}(\Omega)$ itself is treated the same way as $a$. As such, it can also generate its own dissipation function $d_{s-\varsigma}$. The augmented variational principle then yields the conditions
\begin{subequations}
\begin{align}
\label{eq:a}
\langle \delta b, \mathcal{D}_t a \rangle & = d_a \left(1, j_a, \delta b \right), & \forall \;\delta b \in \Lambda^{n-k}(\Omega), \\
\label{eq:s_sigma}
\langle w, \mathcal{D}_t (s - \varsigma) \rangle & = d_{s-\varsigma} \left(1, j_{s-\varsigma}, w \right), & \forall \; w \in \Lambda^0(\Omega), \\
\label{eq:sigma}
\left\langle w, \mathcal{D}_t \varsigma \right\rangle 
& =
d_a \left( \frac{w}{\frac{\delta \ell}{\delta s}}, j_a, \mathcal{D}_t b \right)
+
d_{s-\varsigma} \left( \frac{w}{\frac{\delta \ell}{\delta s}}, j_{s-\varsigma}, \mathcal{D}_t \gamma \right), & \forall \; w \in \Lambda^0(\Omega).
\end{align}
\end{subequations}
Splitting the entropy equation into two contributions cements the analogy between $s-\varsigma$ and $a$. If $d_{s - \varsigma}$ is taken as the continuous dissipation function $d^c$, one recovers the usual heat flux. While $\varsigma$ is the part of the entropy that is {\it produced}, $s-\varsigma$ is the part that is {\it exchanged}. Since $\frac{\delta \ell}{\delta s-\varsigma} = \frac{\delta \ell}{\delta s}$, the function $d_{s-\varsigma}$ and flux $j_{s-\varsigma}$ will be respectively denoted by $d_s$ and $j_s$ in the following. Eliminating the variable $\varsigma$, the resulting equations in weak form read
\begin{subequations}
\begin{align}
\left\langle \frac{\partial }{\partial t}\frac{\delta \ell}{\delta u}, v\right\rangle 
- 
\left\langle \frac{\delta \ell}{\delta u}, \pounds_uv\right\rangle 
& = - 
\left\langle \frac{\delta \ell}{\delta a}, \pounds_va \right\rangle - 
\left\langle \frac{\delta \ell}{\delta s}, \pounds_vs \right\rangle, 
& \forall \;v\in\mathfrak{g}, \\
\left\langle w, \mathcal{D}_t s \right\rangle 
-
d_{s} \left(1, j_{s-\varsigma}, w \right)
& =
d_a \left( \frac{w}{\frac{\delta \ell}{\delta s}}, j_a, \mathcal{D}_t b \right)
+
d_{s} \left( \frac{w}{\frac{\delta \ell}{\delta s}}, j_{s}, \mathcal{D}_t \gamma \right), 
& \forall \; w \in \Lambda^0(\Omega), \\
\label{eq:weak_form_advection_diffusion}
\langle \delta b, \mathcal{D}_t a \rangle 
& = 
d_a \left(1, j_a, \delta b \right), 
& \forall \;\delta b \in \Lambda^{n-k}(\Omega).
\end{align}
\end{subequations}
Note that the variational principle naturally produces two distinct instances of the function $d_s$ in the entropy equation. This corresponds to the classical splitting of an energy flux between an entropy flux and an entropy production.

\begin{remark}[Natural extension]
The variational principle naturally extends to an arbitrary number of variables $(a_i)$. Each variable can give rise to one or several thermodynamic affinities and associated dissipation functions. Some thermodynamic affinities can also depend on a subset of the $(a_i)$, as it is the case, for instance, for chemical reactions. More details will be given in the next section.
\end{remark}

\paragraph{Energy conservation.} The total energy $E$ of the system is defined as
\begin{equation}
E = \left\langle \frac{\delta \ell}{\delta u}, u \right\rangle - \ell.
\end{equation}
Simple computations show that it is conserved over time
\begin{equation}
\label{eq:conservation_energy}
\frac{\text{d}}{\text{d}t} E = 
\left\langle \frac{\partial}{\partial t} \frac{\delta \ell}{\delta u}, u \right\rangle 
- \left\langle \frac{\delta \ell}{\delta a}, \frac{\partial}{\partial t} a \right\rangle 
- \left\langle \frac{\delta \ell}{\delta s}, \frac{\partial}{\partial t} s \right\rangle = 0.
\end{equation}
Consistency with the second principle of thermodynamics depends on the dissipation functions and will be discussed in the next section. In the case where the Lagrangian is of the form \eqref{eq:var1_Omega4}, conservation of energy will also be given a local form.

\subsection{Strong form of the equations}\label{strong_form}

Dissipation is protean and trying to comprehensively list all possible irreversible processes is a wager lost in advance. In the case of differential forms, the two natural closures \eqref{eq:two_fluxes} are here studied, together with their associated strong form.

\subsubsection{Discrete flux}

\paragraph{Formulation.} 
Given two differential forms of same degree and physical nature, a discrete flux can be written to describe exchanges between the two (e.g. a chemical reaction between two particle number density given as volume forms). Given $a_1, a_2 \in \Lambda^k(\Omega)$ be two such forms, let's consider the thermodynamic affinity $\frac{\delta \ell}{\delta a_1} - \frac{\delta \ell}{\delta a_2}$ and associated thermodynamic flux $j(a_1 a_2) \in \Lambda(\Omega)^{n-k}$. The previous (reduced) variational principle \ref{vp:4} is adapted by writing
\begin{equation}
\delta \int_0^T \ell(u(t), a(t), s(t)) + \sum_{i \in \{1,2 \}}\left\langle \mathcal{D}_t b_i (t), a_i(t) \right\rangle + \left\langle \mathcal{D}_t \gamma (t), s(t) - \varsigma(t) \right\rangle {\rm d} t = 0
\end{equation}
together with the phenomenological and variational constraints
\begin{subequations}
\begin{align}
\langle w, \mathcal{D}_t {\varsigma} \rangle & = d^d\left(\frac{w}{\frac{\delta \ell}{\delta s}}, j_{a_1 a_2}, \mathcal{D}_t b_1 - \mathcal{D}_t b_2 \right) = \int_\Omega \frac{w}{\frac{\delta \ell}{\delta s}} \left( \mathcal{D}_t b_1 - \mathcal{D}_t b_2 \right) \wedge j_{a_1 a_2}, \\
\langle w, \mathcal{D}_\delta {\varsigma} \rangle & = d^d\left(\frac{w}{\frac{\delta \ell}{\delta s}}, j_{a_1 a_2}, \mathcal{D}_\delta b_1 - \mathcal{D}_\delta b_2 \right) = \int_\Omega \frac{w}{\frac{\delta \ell}{\delta s}} \left( \mathcal{D}_\delta b_1 - \mathcal{D}_\delta b_2 \right) \wedge j_{a_1 a_2}.
\end{align}
\end{subequations}

\paragraph{Strong form of the equations.} 
It can be shown that the previous adapted variational principle yields the following equations, written in strong form
\begin{subequations}
\begin{align}
\frac{\partial}{\partial t} \frac{\delta \ell}{\delta u} + \pounds_u \frac{\delta \ell}{\delta u} & = \frac{\delta \ell}{\delta a_1} \diamond a_1 + \frac{\delta \ell}{\delta a_2} \diamond a_2 + \frac{\delta \ell}{\delta s} \diamond s, \\
\mathcal{D}_t s & = \frac{1}{-\frac{\delta \ell}{\delta s}} \left( \frac{\delta \ell}{\delta a_1} - \frac{\delta \ell}{\delta a_2} \right) \wedge j_{a_1 a_2}, \\
\mathcal{D}_t a_1 & = + j_{a_1 a_2}, \\
\mathcal{D}_t a_2 & = - j_{a_1 a_2}.
\end{align}
\end{subequations}
As expected, the formulation creates a discrete flux $j_{a_1 a_2}$ between the two quantities, and the following conservation law is immediately recovered as a consequence
\begin{equation}
\label{eq:conservation_a1_a2}
\mathcal{D}_t (a_1 + a_2) = 0.
\end{equation}

\paragraph{Generalization.}
This example extends to a set of more than two variables $\{a_i\}_{1 \leq i \leq N} \in (\Lambda^k(\Omega))^N$ and to a thermodynamical affinity given as an arbitrary linear combination of their dual variables $\{\frac{\delta \ell}{\delta a_i}\}_{1 \leq i \leq N} \in (\Lambda^{n-k}(\Omega))^N$. This is for instance relevant for chemical reactions involving an arbitrary number of materials, and with arbitrary stoichiometry coefficients. In practice, let $\{a_i\}_{1 \leq i \leq N} \in \mathbb{R}^N$ and consider a physical thermodynamical process $\alpha$ whose thermodynamical affinity is $x^\alpha = \sum_i \lambda_i^\alpha \frac{\delta \ell}{\delta a_i}$ and whose thermodynamical flux is written as $j_\alpha$. The associated (reduced) phenomenological constraint is then
\begin{equation}
\langle w, \mathcal{D}_t \varsigma \rangle = \int_\Omega \frac{w}{\frac{\delta \ell}{\delta s}} \left( \sum_{i=1}^N \lambda_i^\alpha \mathcal{D}_t b_i\right) \wedge j_\alpha.
\end{equation}
and likewise for the variational constraint. This in turn gives the following $N$ evolution equations
\begin{equation}\label{a_i_equations}
\mathcal{D}_t a_i = \lambda_i^\alpha j_\alpha, \quad \forall 1 \leq i \leq N.
\end{equation}
This system gives rise to $N-1$ independent conservation laws of the form $\mathcal{D}_t (\mu_\alpha^i a_i) = 0$, where $\sum_i \mu_\alpha^i \lambda_i^\alpha = 0$.

\begin{remark}
When considering several thermodynamical processes $\{ \alpha \in \mathcal{A} \}$, it is possible to rewrite the entropy production either as a sum on the processes or a sum on the variables
\begin{equation}
\sum_{\alpha \in \mathcal{A}} \int_\Omega \frac{w}{\frac{\delta \ell}{\delta s}} x^\alpha \wedge j_\alpha 
= \sum_{\alpha \in \mathcal{A}} \int_\Omega \frac{w}{\frac{\delta \ell}{\delta s}} \left( \sum_{i=1}^N \lambda_i^\alpha \frac{\delta \ell}{\delta a_i} \right) \wedge j_\alpha
= 
\sum_{i=1}^N \int_\Omega \frac{w}{\frac{\delta \ell}{\delta s}} \frac{\delta \ell}{\delta a_i} \wedge \left( \sum_{\alpha \in \mathcal{A}} \lambda_i^\alpha  j_\alpha \right)
=
\sum_{i=1}^N \int_\Omega \frac{w}{\frac{\delta \ell}{\delta s}} \frac{\delta \ell}{\delta a_i} \wedge j_{a_i}.
\end{equation}
The first point of view is phenomenologically driven as it isolates individual physical processes and automatically guarantees that conservation law holds. The second point of view allows for an easier algebraic treatment of the equations but requires \textit{a posteriori} constraints on the fluxes $j_{a_i}$ to satisfy conservation laws.
\end{remark}

\paragraph{Local energy conservation} If the Lagrangian is of the form \eqref{eq:var1_Omega4}, then the total energy density $\mathfrak{e} = \frac{\partial \mathfrak{l}}{\partial u} \cdot u - \mathfrak{l}$ satisfies the following local conservation law
\begin{equation}
\label{eq:local_conservation_energy_1}
\mathcal{D}_t \mathfrak{e} = \text{d} \left[ \frac{\partial \mathfrak{l}}{\partial s} \wedge i_u s + (-1)^{n-k} \frac{\partial \mathfrak{l}}{\partial a} \wedge i_u a - i_u \mathfrak{l} \right].
\end{equation}
It is a conservation law as it can be integrated on any time-varying volume $\mathcal{V}(t)$; the integrated value of $\mathfrak{e}$ is then function of fluxes at the boundary $\partial \mathcal{V}(t)$ thanks to Stoke's theorem. Note that if $a$ is a volume form (i.e. $k=n$), then \eqref{eq:local_conservation_energy_1} can alternatively be written
\begin{equation}
\mathcal{D}_t \mathfrak{e} = \pounds_u \left( \frac{\partial \mathfrak{l}}{\partial s} s + \frac{\partial \mathfrak{l}}{\partial a} a - \mathfrak{l} \right).
\end{equation}

\subsubsection{Continuous flux}

\label{sec:continuous_flux}

\paragraph{Formulation.}
Given a differential form $a \in \Lambda(\Omega)^k$, the exterior derivative of its dual variable $\text{d} \frac{\delta \ell}{\delta a} \in \Lambda(\Omega)^{n-k+1}$ is a natural choice for a thermodynamic affinity. Its associated flux $j_a \in \Lambda(\Omega)^{k-1}$ can describes \textit{spatial} exchanges such as heat flux or resistivity in MHD. With this choice, the phenomenological and variation constraints of the (reduced) variational principle \ref{vp:4} can now be explicitely written as
\begin{subequations}
\begin{align}
\langle w, \mathcal{D}_t \varsigma \rangle & = d^c \left( \frac{w}{\frac{\delta \ell}{\delta s}}, j_a, \mathcal{D}_t b \right) = \int_\Omega \frac{w}{\frac{\delta \ell}{\delta s}} \left( \text{d} \mathcal{D}_t b \right) \wedge j_a, \\
\langle w, \mathcal{D}_\delta \varsigma \rangle & = d^c \left( \frac{w}{\frac{\delta \ell}{\delta s}}, j_a, \mathcal{D}_\delta b \right) = \int_\Omega \frac{w}{\frac{\delta \ell}{\delta s}} \left( \text{d} \mathcal{D}_\delta b \right) \wedge j_a.
\end{align}
\end{subequations}

\paragraph{Strong form of the equations.} The weak equation \eqref{eq:weak_form_advection_diffusion} describing the evolution of $a$ now becomes, for all $\delta b \in \Lambda(\Omega)^{n-k}$,
\begin{equation}
\langle \delta b, \mathcal{D}_t a \rangle = \int_\Omega \text{d} \delta b \wedge j_a = \int_\Omega \text{d} \left( \delta b \wedge j_a \right) + (-1)^{n-k+1} \delta b \wedge \text{d} j_a .
\end{equation}
Hence, the strong form of the equations is
\begin{subequations}
\begin{align}
\frac{\partial}{\partial t} \frac{\delta \ell}{\delta u} + \pounds_u \frac{\delta \ell}{\delta u} & = \frac{\delta \ell}{\delta a} \diamond a + \frac{\delta \ell}{\delta s} \diamond s, \\
\mathcal{D}_t s & = \frac{1}{-\frac{\delta \ell}{\delta s}} \left( \text{d} \frac{\delta \ell}{\delta a} \right) \wedge j_{a}, \\
\label{eq:strong_form_advection_diffusion}
\mathcal{D}_t a & = (-1)^{n-k+1} \text{d} j_a,
\end{align}
\end{subequations}
supplemented by the following boundary conditions
\begin{equation}
\label{eq:homogeneous_BC}
i_{\partial \Omega \rightarrow \Omega}^* j_a = 0,
\end{equation}
where $i_{\partial \Omega \rightarrow \Omega} : \partial \Omega \rightarrow \Omega$ is the canonical inclusion. 

\paragraph{Conservation laws.} When it is integrated over a submanifold following the flow of the fluid $\mathcal{N}(t) = \varphi(t)(\mathcal{N}_0) \subset \Omega$ of dimension $k$, with boundary $\partial \mathcal{N}(t)$, equation \eqref{eq:strong_form_advection_diffusion} becomes
\begin{equation}
\label{eq:conservation_a}
\frac{\rm d}{{\rm d} t} \int_{\mathcal{N}(t)} i^{*}_{\mathcal{N}(t) \rightarrow \Omega} a = (-1)^{n-k+1} \int_{\partial \mathcal{N}(t)} i^{*}_{\partial \mathcal{N}(t) \rightarrow \Omega} j_a.
\end{equation}
It is a conservation equation as the total integrated value only varies through fluxes at the boundary. In particular, it is constant if $\partial \mathcal{N} = \varnothing$ or $\partial \mathcal{N} \subset \partial \Omega$ (because of \eqref{eq:homogeneous_BC}). The flux $j_a$ also have an effect on the local conservation of energy, which now reads
\begin{equation}
\label{eq:local_conservation_energy_2}
\mathcal{D}_t \mathfrak{e} = \text{d} \left[ \frac{\partial \mathfrak{l}}{\partial s} \wedge i_u s + (-1)^{n-k} \frac{\partial \mathfrak{l}}{\partial a} \wedge i_u a - i_u \mathfrak{l} + \frac{\partial \mathfrak{l}}{\partial a} \wedge j_a \right].
\end{equation}

\paragraph{Other boundary conditions.} Dirichlet and non-homogeneous Neumann conditions can be incorporated into the variational principle by changing the variational constraints
\begin{equation}
\langle w, \mathcal{D}_\delta \varsigma \rangle = \int_\Omega \frac{w}{\frac{\delta \ell}{\delta s}} (\text{d} \mathcal{D}_\delta b) \wedge j_a - \int_{\partial \Omega} \frac{w}{\frac{\delta \ell}{\delta s}} i^*_{\partial \Omega \rightarrow \Omega} \left( \mathcal{D}_\delta b \wedge j_a \right),
\end{equation}
while the phenomenological constraint is respectively replaced by
\begin{subequations}
\begin{align}
\langle w, \mathcal{D}_t \varsigma \rangle & = \int_\Omega \frac{w}{\frac{\delta \ell}{\delta s}} (\text{d} \mathcal{D}_t b) \wedge j_a - \int_{\partial \Omega} \frac{w}{\frac{\delta \ell}{\delta s}} i^*_{\partial \Omega \rightarrow \Omega} \left( (\mathcal{D}_t b + c_0) \wedge j_a \right), \\
\langle w, \mathcal{D}_t \varsigma \rangle & = \int_\Omega \frac{w}{\frac{\delta \ell}{\delta s}} (\text{d} \mathcal{D}_t b) \wedge j_a - \int_{\partial \Omega} \frac{w}{\frac{\delta \ell}{\delta s}} i^*_{\partial \Omega \rightarrow \Omega} \left( \mathcal{D}_t b \wedge (j_a - j_0) \right).
\end{align}
\end{subequations}
In the first case, equation \eqref{eq:homogeneous_BC} is replaced with the Dirichlet boundary condition $i_{\partial \Omega \rightarrow \Omega}^* (\frac{\delta \ell}{\delta a} - c_0) = 0$. In the second case, equation \eqref{eq:homogeneous_BC} is replaced with the Neumann boundary condition $i_{\partial \Omega \rightarrow \Omega}^* (j_a - j_0) = 0$. This consistently extends to $k$-forms the situation considered in \cite{Gawlik2024} for thermal boundary conditions.

\section{Flux closures and the second law of thermodynamics}
\label{sec:dissipation}

Compliance of the equations with the second law of thermodynamics is now studied. Flux closures are then given and include couplings between thermodynamical affinities. Onsager's principle and Curie's symmetry principle are both discussed. Finally, a small comment is made on how to include viscosity in the formulation.

\subsection{General flux closures}

\paragraph{Individual process.} 
Given a thermodynamical process with affinity $x^\alpha \in \Lambda^k(\Omega)$ and flux $j_\alpha \in \Lambda^{n-k}(\Omega)$, the entropy production is
\begin{equation}
\mathcal{D}_t s = \frac{1}{-\frac{\delta \ell}{\delta s}} x^\alpha \wedge j_\alpha.
\end{equation}
By virtue of the second law of thermodynamics, it must be positive; note that positivity can only be understood if an orientation of the manifold is given. Assuming the temperature $T = -\frac{\delta \ell}{\delta s}$ is positive (which is the case for fluid applications), this leads to the following local condition
\begin{equation}
(x^\alpha \wedge j_\alpha)(e_1, \cdots, e_n) \geq 0,
\end{equation}
for one (or equivalently, every) positively oriented basis $(e_i)$ of the tangent space at every point of the manifold. For states that are near thermodynamical equilibrium, fluxes can reasonably be expressed as linear functions of forces \cite{Groot1962}
\begin{equation}
\label{eq:indiv_flux_closure}
j_\alpha = \mathcal{K}_\alpha x^\alpha,
\end{equation}
where $\mathcal{K}_\alpha: \Lambda^k(\Omega) \rightarrow \Lambda^{n-k}(\Omega)$ is an operator that depends locally on the fluid state. In order to comply with the second law of thermodynamics, the operator must be so that, for a positively oriented basis,
\begin{equation}
\left( x^\alpha \wedge j_\alpha \right) (e_1, \cdots, e_n)
=
\left( x^\alpha \wedge \mathcal{K}_\alpha x^\alpha \right) (e_1, \cdots, e_n)
\geq 0.
\end{equation}
Given a metric and the associated Hodge star operator $\star$, $\mathcal{K}_{\alpha}$ can be factorized into $\mathcal{K}_{\alpha} = \star K_\alpha$. In this case, the operator $K_\alpha: \Lambda^k(\Omega) \rightarrow \Lambda^k(\Omega)$ must be positive semi-definite. Note however that the metric only gives a representation of the operator $\mathcal{K}_{\alpha}$ and the covariant expression $x^\alpha \wedge j_\alpha$; the metric is completely arbitrary and does not even need to be the physical one.

\begin{remark}
This closure applies to discrete ($x^\alpha = \sum_i \lambda^\alpha_i \frac{\delta \ell}{\delta a_i}$) and continuous ($x^\alpha = \text{d} \frac{\delta \ell}{\delta a}$) affinities alike.
\end{remark}

\paragraph{Cross-effects and Onsager's principle.} When several thermodynamical processes are considered, the second law of thermodynamics applies to the sum of their contribution
\begin{equation}
\sum_{\alpha \in \mathcal{A}} (x^\alpha \wedge j_\alpha) (e_1, \cdots, e_n) \geq 0.
\end{equation}
Imposing the previous closure \eqref{eq:indiv_flux_closure} for each flux individually is sufficient to ensure positive entropy production; however, it is not necessary, since fluxes may couple different processes. In the most general case, cross-effects can be written between each pair of processes
\begin{equation}
j_\alpha = \sum_{\beta \in \mathcal{A}} \mathcal{K}_{\alpha \beta} x^\beta,
\end{equation}
where $\mathcal{K}_{\alpha \beta}: \Lambda^{\text{deg}(x^{\beta})}(\Omega) \rightarrow \Lambda^{n-\text{deg}(x^{\alpha})}(\Omega)$ is a linear operator. Writing $\mathcal{K}^T_{\alpha \beta}: \Lambda^{\text{deg}(x^{\alpha})}(\Omega) \rightarrow \Lambda^{n-\text{deg}(x^{\beta})}(\Omega)$ for the adjoint operator (with respect to the wedge product), we may write
\begin{subequations}
\label{eq:entropy_production_computation}
\begin{align}
\sum_{\alpha \in \mathcal{A}} x^\alpha \wedge j_\alpha & = \sum_{\alpha, \beta \in \mathcal{A}} x^\alpha \wedge (\mathcal{K}_{\alpha \beta} x^\beta) \\
& = \sum_{\alpha \in \mathcal{A}} x^\alpha \wedge (\mathcal{K}_{\alpha \alpha} x^\alpha) + \sum_{\{\alpha, \beta \} \subset \mathcal{A}} \left[ x^\alpha \wedge ( \mathcal{K}_{\alpha \beta} x^\beta ) +  x^\beta \wedge ( \mathcal{K}_{\beta \alpha} x^\alpha ) \right] \\
& = \sum_{\alpha \in \mathcal{A}} x^\alpha \wedge (\mathcal{K}_{\alpha \alpha} x^\alpha) + \sum_{\{ \alpha, \beta \} \subset \mathcal{A}} \left[ \left( \mathcal{K}_{\beta \alpha} + \mathcal{K}^T_{\alpha \beta} \right) x^\alpha \right] \wedge x^\beta.
\end{align}
\end{subequations}
The Onsager--Casimir principle states that the set $\mathcal{A}$ can be split into two disjoint subsets $\mathcal{A} = \mathcal{A}_+ \cup \mathcal{A}_-$ such that
\begin{subequations}
\begin{align}
\label{eq:onsager_symm}
\mathcal{K}^T_{\alpha \beta} & = \mathcal{K}_{\beta \alpha} & \text{if} \quad \alpha,\beta \in \mathcal{A}_+ \quad \text{or} \quad  \alpha,\beta \in \mathcal{A}_-,  \\
\label{eq:onsager_skewsymm}
\mathcal{K}^T_{\alpha \beta} & = -\mathcal{K}_{\beta \alpha} & \text{otherwise}.
\end{align}
\end{subequations}
In particular, $\mathcal{K}^T_{\alpha \alpha} = \mathcal{K}_{\alpha \alpha}$ so it is a symmetric operator. The subsets $\mathcal{A}_+$ and $\mathcal{A}_-$ are related to the effect of time-reversibility on their underlying thermodynamical affinities: that of $\mathcal{A}_+$ are invariant while that of $\mathcal{A}_-$ take on a minus sign after time reversibility. In this case, \eqref{eq:entropy_production_computation} becomes
\begin{equation}
\sum_{\alpha \in \mathcal{A}} x^\alpha \wedge j_\alpha 
=
\sum_{\alpha, \beta \in \mathcal{A}_+} x^\alpha \wedge (\mathcal{K}_{\alpha \beta} x^\beta)
+
\sum_{\alpha, \beta \in \mathcal{A}_-} x^\alpha \wedge (\mathcal{K}_{\alpha \beta} x^\beta).
\end{equation}
The mixed operators $\mathcal{K}_{\alpha \beta}$ with $\alpha \in \mathcal{A}_\pm$ and  $\beta \in \mathcal{A}_\mp$ do not contribute to the entropy equation and are thus not subject to any additional condition beyond the Onsager symmetry \eqref{eq:onsager_skewsymm}. As for the sets of operators $\mathcal{K}_{\alpha \beta}$ with $\alpha, \beta \in \mathcal{A}_+$, they must be so that the first sum of the right-hand side is positive; likewise for the operators with $\alpha, \beta \in \mathcal{A}_-$ and the second sum.

\subsection{Curie's principle}

In the context of thermodynamics, Curie's principle states that if the system possesses some intrinsic symmetry, then dissipation must also satisfy such symmetry. In the case where the system is isotropic, the statement is often simplified into the impossibility of cross-effects between quantities of different tensorial natures; in the present case, $\mathcal{K}_{\alpha \beta} = 0$ if the thermodynamical affinities associated to $\alpha$ and $\beta$ are differential forms of different degrees. This simplified version is however a non-trivial consequence of the first statement, and can be given a more accurate formulation that actually allows for \textit{some} cross-effect between variables of different geometrical natures. This formulation uses the language of group theory and can be generalized to systems with different type of invariance.

\subsubsection{Curie's principle for general symmetry}

The set of symmetries of a system forms a group $H$. This group may be continuous (e.g., the isometry group $O(n)$ for isotropic fluids) or discrete (e.g., the invariance group of a crystal structure). The group acts on thermodynamic affinities and thermodynamic fluxes. The present work focuses on the case in which the group acts separately on each (finite-dimensional) fiber. A local formulation, as opposed to one using global symmetries of the domain, is favored because invariance is assumed to come from the fluid properties; it should thus not depend on the domain it is embedded in. The action on each fiber is assumed to be linear and therefore endows the fiber spaces with the structure of group representations. Some basic elements of representation theory are recalled here; a thorough exposition of the subject can be found in \cite{Knapp2001, Zee2016}.

\paragraph{Representations of groups.} 
Let $H$ be a group. A (left) representation of a vector space $V$ is a group morphism $\phi: H \rightarrow GL(V)$; that is a function satisfying, for all $h_1, h_2 \in H$,
\begin{equation}
\phi(h_1 h_2) = \phi(h_1) \circ \phi(h_2).
\end{equation}
The function $\phi$ is often omitted, and the action is written directly as $\phi(h)(v) = h \cdot v$. Subrepresentations are vector subspaces $W \subset V$ that are stable under the action of the group, i.e. $h\cdot w \in W$ for all $w \in W$ and $h \in H$. An irreducible representation is a representation that has no subrepresentations other than the zero subspace and the whole space $V$. When $H$ is finite or compact, every finite-dimensional representation is completely reducible, meaning that $V$ can be written as a direct sum of irreducible subrepresentations $W_i$. In particular, if the representation is not irreducible, it decomposes into several irreducible components.

Let $x^\alpha \in \Lambda^k (\Omega)$ be a thermodynamic affinity, and $j_\alpha \in \Lambda^{n-k} (\Omega)$ its corresponding flux. Given a representation of $H$ on the fiber space $V = \Lambda^{n-k}_x \Omega$, it naturally defines a representation on the space $V^* = \Lambda^k_x \Omega$ that preserves the density-valued dual pairing
\begin{equation}
\label{eq:dual_action}
(h \cdot_* x) \wedge (h \cdot j) = x \wedge j, \quad \forall \;(x,j) \in V^* \times V.
\end{equation}
Note that the space of $k$-alternate forms may already be given a natural structure of representation whose action $h \cdot x$ may not coincide with the dual action $h \cdot_* x$. If the representation on $V$ splits into several irreducible components, it means that $x$ and $j$ can be written as the sum of several elementary affinities. This is typically what happens when the viscous stress tensor is split into its shear and bulk parts for isotropic fluids (more on that in §\ref{sec:viscosity}).

\paragraph{Intertwining operators.} Intertwining operators, or representation morphisms, are linear operators that commute with the group action. Let $V_1, V_2$ be two representations of the same group $H$. Then $\mathcal{K}: V_1 \rightarrow V_2$ is an intertwining operator if it is linear and satisfies
\begin{equation}
\mathcal{K}(h \cdot v) = h \cdot \mathcal{K}(v), \quad \forall \; v \in V_1, \quad \forall \; h \in H.
\end{equation}
Note that the group action on $v \in V_1$ and $\mathcal{K}(v) \in V_2$ is not necessarily the same if $V_1 \neq V_2$. The space of intertwining operators is written $\text{Hom}_H(V_1, V_2)$. Two representations are said to be isomorphic if there exists an invertible intertwining operator between the two.

As before, let $\mathcal{A}$ be the set of dissipative processes. The group $H$ is assumed to act on the fiber $V^\alpha = \Lambda^{\text{deg}(x^\alpha)}_x \Omega$ of the space of all thermodynamic affinities $x^\alpha$; the fiber space $V_\alpha = \Lambda^{n-\text{deg}(x^\alpha)}_x \Omega$ of thermodynamic fluxes $j_\alpha$ is endowed with the corresponding dual action \eqref{eq:dual_action}. Invariance of the entropy equation under the action of $H$ can be locally written as
\begin{equation}
\sum_{\alpha, \beta \in \mathcal{A}} \left( \mathcal{K}_{\alpha \beta} (h \cdot x^\beta) \right) \wedge (h \cdot x^\alpha) = \sum_{\alpha, \beta \in \mathcal{A}} (\mathcal{K}_{\alpha \beta} x^\beta) \wedge x^\alpha.
\end{equation}
Because of \eqref{eq:dual_action}, this leads to the natural condition
\begin{equation}
\mathcal{K}_{\alpha \beta} (h \cdot x^\beta) = h \cdot_* (\mathcal{K}_{\alpha \beta} x^\beta), \quad \forall\; x^\beta \in V^\beta, \quad \forall\; h \in H.
\end{equation}
Then $\mathcal{K}_{\alpha \beta}$ commutes with the action of $H$ and thus belongs to $\operatorname{Hom}_H(V^\beta, V_\alpha)$. This is exactly the condition for $\mathcal{K}_{\alpha \beta}$ to be consistent with the symmetries of the system. It is thus essential to understand the structure of $\operatorname{Hom}_H(V^\beta, V_\alpha)$ to completely characterize flux closures. Schur's lemma is a crucial result in this endeavor.

\begin{lemma}[Schur's lemma]
Let $V_1$ and $V_2$ be two irreducible representations, then either 
\begin{itemize}
\item[\textit{(i)}] the two representations are not isomorphic and $\operatorname{Hom}_H(V_1, V_2) = \{ 0 \}$,
\item[\textit{(ii)}] the two representations are isomorphic and $\operatorname{Hom}_H(V_1, V_2)$ is a non-trivial division algebra over $\mathbb{R}$.
\end{itemize}
Besides, according to Frobenius theorem, the only finite-dimensional associative division algebras over $\mathbb{R}$, up to isomorphism, are $\mathbb{R}$ itself, the complex numbers $\mathbb{C}$ and the space of quaternions $\mathbb{H}$.
\end{lemma}

\paragraph{Classification of intertwining operators.} Let $V_1$ and $V_2$ be two representations and assume they split into pairwise non-isomorphic irreducible components $\{W^i\}_i$,
\begin{equation}
\label{eq:rep}
V_1 \approx \bigoplus_i (W^i)^{n_i}, \quad \quad \quad V_2 \approx \bigoplus_i (W^i)^{n_i}.
\end{equation}
Such a decomposition always exists if $H$ is finite or compact. Then the space of intertwining operators $\operatorname{Hom}_H(V_1, V_2)$ is isomorphic, as a division algebra, to
\begin{subequations}
\label{eq:classification_intertwining_operators}
\begin{align}
\operatorname{Hom}_H(V_1, V_2) & \approx \bigoplus_{i,j} \operatorname{Hom}_H((W^i)^{n_i}, (W^j)^{m_j}), \\
& \approx \bigoplus_{i,j} \text{Mat}_{n_i \times m_j}(\operatorname{Hom}_H(W^i, W^j)), \\
& \approx \bigoplus_{i} \text{Mat}_{n_i \times m_j}(D_i),
\end{align}
\end{subequations}
where $D_i = \operatorname{Hom}_H(W^i, W^i)$ is either $\mathbb{R}, \mathbb{C}$ or $\mathbb{H}$. This results simply states that, in a basis adapted to decompositions \eqref{eq:rep}, the matrix representation of intertwining operators can be decomposed into blocks where each block is an element of $D_i$ if the two underlying irreducible components are isomorphic, or zero if they are not.

\subsubsection{Isotropy as invariance under the orthogonal group}

The previous point is illustrated by the physically relevant example of isotropy. An isotropic system behaves identically in all directions, which can be reformulated as a local invariance principle under the action of the orthogonal group $O(n)$, generated by rotations and reflections.

\paragraph{Action of the group $O(n)$ on alternate forms.} Given a metric $g$ on $\Omega$, $O(n)$ is \textit{locally} identified with the group of linear operators $R$ that preserves the inner product on $T_x \Omega$, that is, for all $u,v \in T_x \Omega$
\begin{equation}
\label{eq:O(n)_vectors}
g(R \cdot u, R \cdot v) = g(u, v).
\end{equation}
From this, it is possible to define two metric-preserving representations $V^k = \Lambda^k_x \Omega$ by defining, for all vectors $u_1, \cdots, u_k \in T_x \Omega$
\begin{subequations}
\begin{align}
\label{eq:O(n)_forms_1}
R \triangleright w(u_1, \cdots, u_k) & = w(R^{-1} \cdot u_1, \cdots, R^{-1} \cdot u_k), \\
\label{eq:O(n)_forms_2}
R \triangleleft w(u_1, \cdots, u_k) & = w(R^{-1} \cdot u_1, \cdots, R^{-1} \cdot u_k) \times \text{det}(R).
\end{align}
\end{subequations}
The two distinct representations will respectively be written $V^k_{\triangleright}$ and $V^k_{\triangleleft}$ to avoid any ambiguity. Note that both can be recovered from the other one by tensorialization with the one-dimensional involutive representation $\text{det} : O(n) \rightarrow \mathbb{R} = GL(\mathbb{R})$. An important connection between the two is that together they preserve the wedge product
\begin{equation}
(R \triangleright x) \wedge (R \triangleleft j) = (R \triangleleft x) \wedge (R \triangleright j) = x \wedge j.
\end{equation}
This means that the dual representation of $V^k_{\triangleright}$ as defined in \eqref{eq:dual_action}, is $V^{n-k}_{\triangleleft}$. This is natural as $O(n)$ preserves the metric but not the present density-valued pairing. The Hodge star operator then acts as the connection between the two duals by
\begin{equation}
(R \triangleright x) \wedge (R \triangleleft \star y) = x \wedge \star y = g(x, y) \text{d}x = g(R \triangleright x, R \triangleright y) \text{d}x = (R \triangleright x) \wedge (\star R \triangleright y),
\end{equation}
so that $R \triangleright \star = \star R \triangleleft$ for all $R \in O(n)$; for the same reason, $R \triangleleft \star = \star R \triangleright$. Said otherwise, the Hodge star operator is an intertwining operator between $V^k_{\triangleright}$ and $V^{n-k}_{\triangleleft}$. In particular, $V^k_{\triangleright}$ and $V^{n-k}_{\triangleleft}$ are isomorphic as representations. 

\begin{remark}
In practice, choosing the action $\triangleright$ or $\triangleleft$ depends on the physical nature of the thermodynamical affinity and how it behaves under reflections. Examples will be given in §\ref{sec:case_study}.
\end{remark}

\paragraph{Irreducibility and intertwining operators.} It turns out that $V^k_{\triangleright}$ are irreducible representations, for all degrees $k$ \cite[\S 19]{Fulton2013}. Moreover, they are all of the real type and pairwise non-isomorphic. As a result,
\begin{subequations}
\begin{align}
\text{Hom}_H(V^k_\triangleright, V^\ell_\triangleright) 
= 
\text{Hom}_H(V^k_\triangleleft, V^\ell_\triangleleft) 
& =
\left\{ 
\begin{array}{ll}
\text{span}\left( \left\{ \mathbb{I} \right\} \right) & \text{if $\ell = k$}, \\
\text{span}\left( \left\{ 0 \right\} \right) & \text{else}.
\end{array}
\right.
\\
\text{Hom}_H(V^k_\triangleright, V^\ell_\triangleleft) 
=
\text{Hom}_H(V^k_\triangleleft, V^\ell_\triangleright) 
& =
\left\{ 
\begin{array}{ll}
\text{span}\left( \left\{ \star \right\} \right) & \text{if $\ell = n-k$}, \\
\text{span}\left( \left\{ 0 \right\} \right) & \text{else}.
\end{array}
\right.
\end{align}
\end{subequations}
Coming back to Curie's principle and thermodynamics, this has two consequences. First, cross-effects can only occur between thermodynamical forces $x^\alpha$ and $x^\beta$ of respective degrees $k$ and $k$ (if the actions are the same), or $k$ and $n-k$ (if the actions are dual of one another); it thus circles back to the usual statement of Curie's principle. Second, on every fiber, the operator $\mathcal{K}_{\alpha \beta}$ essentially acts as a scalar coefficient. In particular, for an isolated thermodynamic force $x^\alpha$, $\mathcal{K}_{\alpha} = \kappa_\alpha \star$ with $\kappa_{\alpha} \in \mathbb{R}$. In this case,
\begin{equation}
x^\alpha \wedge j_\alpha = x^\alpha \wedge (\star \kappa_\alpha x^\alpha) = \kappa_\alpha g(x^\alpha, x^\alpha) \text{d} x.
\end{equation}
Consistency with the second principle of thermodynamics then requires $\kappa_\alpha \geq 0$.

\begin{remark}[$SO(n)$ symmetry]
A weaker form of isotropy is that related to the sub-group $SO(n) \subset O(n)$ of positive isometries. It is relevant for systems which possess an inner chirality and that are invariant under rotations but not reflections (e.g. fluid in an external magnetic field). Note that the two representations $V^k_\triangleright$ and $V^k_\triangleleft$ coincides in this case. The classification of their intertwining operators is slightly more involved in the case where $2k = n$; details may be found in \cite[\S 19]{Fulton2013}.
\end{remark}

\subsection{Viscosity} 

\label{sec:viscosity}

Although the case where $a$ is a general tensor-like object is outside the scope of this article, this section explains how viscosity can be included in the variational principle. Following the work of \cite{Kanzo2007}, it is actually possible to include it in a manner that is formally equivalent to a continuous flux for differential forms \eqref{eq:continuous_flux}. A metric is however necessary to write the associated momentum flux. 

\paragraph{Formalism.} For that purpose, the viscous tensor is seen as an element of $W = \Gamma(T^* \Omega \otimes \Lambda^{n-1} T ^*\Omega)$, the space of sections of the bundle $T^*\Omega \otimes \Lambda^{n-1} T^*\Omega$. The dual space $W^*$ is identified with the space $\Gamma(T \Omega \otimes T^*\Omega)$ of sections of the bundle $T\Omega \times T^* \Omega$. For $(\mathcal{T}, \mathcal{S}) \in W \times W^*$, the dual pairing reads
\begin{equation}
\langle \mathcal{S}, \mathcal{T} \rangle = \int_\Omega \mathcal{S} \ \dot{\wedge} \ \mathcal{T},
\end{equation}
where $\mathcal{S} \ \dot{\wedge} \ \mathcal{T}$ is the $n-$form obtained by applying the natural pairing on their respective first component and applying the wedge product on their respective second component. More explicitly, it is linearly generated by the rule $(a \otimes b) \ \dot{\wedge} \ (c \otimes d) = c(a) (b \wedge d)$ where $a \in \mathfrak{X}(\Omega)$, $b,c \in \Lambda^1(\Omega)$ and $d \in \Lambda^{n-1}(\Omega)$. 
In coordinates, one writes
\begin{subequations}
\begin{align}
\mathcal{T} & = T^j_i \text{d} x^i \otimes {\rm d}^{n-1} x_j, \\
\mathcal{S} & = S^i_j \partial_{x^i} \otimes {\rm d} x^j,
\end{align}
\end{subequations}
with ${\rm d}^{n-1} x_j= {\rm i}_{\partial_{x^j}}{\rm d} x$, and $\mathcal{S} \ \dot{\wedge} \ \mathcal{T}$ simply reads
\begin{equation}
\mathcal{S} \ \dot{\wedge} \ \mathcal{T} = S^i_j T^j_i \text{d} x.
\end{equation}
Given a vector field $u \in \mathfrak{X}(\Omega)$, its covariant derivative $\nabla^g u$ is seen an element of $W^*$, with local expression $\nabla^g u = \nabla_j v^i \partial_{x^i} \otimes \text{d} x^j$. This allows to define a new differentiation operation $\mathfrak{d}$. It takes an element of $W$ and returns a co-vector density, i.e. a section of $T^*\Omega \otimes \Lambda^n T ^*\Omega$. It is analog to the exterior derivative for differential forms and is defined as satisfying, for every vector field $u \in \mathfrak{X}(\Omega)$,
\begin{equation}
\label{eq:IPP_tensors}
u \cdot_1 \mathfrak{d} \mathcal{T} = \text{d} ( u \cdot_1 \mathcal{T} ) - \nabla^g u \ \dot \wedge \ \mathcal{T},
\end{equation}
where $\cdot_1$ simply denotes the natural product with the first component of $\mathcal{T}$. It can be readily checked that this is a well-defined metric-dependent operator \cite{Kanzo2007}. In coordinates, one gets
\begin{equation}
\mathfrak{d}\mathcal{T}= ( \partial _{x^j}\mathcal{T}^j_i - \Gamma^k_{ji}\mathcal{T}^j_k) {\rm d}x^i \otimes {\rm d}^n x.
\end{equation}

\paragraph{Variational principle with viscosity.} For viscous applications, the configuration group $G$ is usually replaced with $G_0 = \operatorname{Diff}_0(\Omega)$, the group of diffeomorphisms that fix $\partial \Omega$ pointwise. Its Lie algebra $\mathfrak{g}_0$ is the space of vector fields that vanish on $\partial \Omega$. This is a usual assumption and will conveniently allow for the cancellation of the boundary term in \eqref{eq:IPP_tensors}. The viscous Eulerian dissipation function $d_{\frac{\delta \ell}{\delta u}} : \Lambda^0(\Omega) \times W \times \mathfrak{g}_0 \rightarrow \mathbb{R}$ is defined as
\begin{equation}
d_{\frac{\delta \ell}{\delta u}} \left( \frac{w}{\frac{\delta \ell}{\delta s}}, j_{\frac{\delta \ell}{\delta u}}, u \right) = \int_\Omega \frac{w}{\frac{\delta \ell}{\delta s}} \nabla^g u \ \dot{\wedge} \ j_{\frac{\delta l}{\delta u}},
\end{equation}
with $j_{\frac{\delta \ell}{\delta u}}\in W$ representing the viscous stress tensor. If the phenomenological and variational constraints \eqref{eq:eulerian_phenomemo_constraint}-\eqref{eq:eulerian_variational_constraint} are now written
\begin{subequations}
\begin{align}
\left\langle w, \mathcal{D}_t \varsigma \right\rangle 
& =
d_{\frac{\delta \ell}{\delta u}} \left( \frac{w}{\frac{\delta \ell}{\delta s}}, j_{\frac{\delta l}{\delta u}}, u \right), \\
\left\langle w, \mathcal{D}_\delta \varsigma \right\rangle
& = 
d_{\frac{\delta \ell}{\delta u}} \left( \frac{w}{\frac{\delta \ell}{\delta s}}, j_{\frac{\delta \ell}{\delta u}}, v \right),
\end{align}
\end{subequations}
it leads to the following weak system of equations:
\begin{subequations}
\begin{align}
\left\langle \frac{\partial }{\partial t}\frac{\delta \ell}{\delta u}, v\right\rangle 
- \left\langle \frac{\delta \ell}{\delta u}, \pounds_uv\right\rangle 
+ \left\langle \frac{\delta \ell}{\delta a}, \pounds_va \right\rangle
+ \left\langle \frac{\delta \ell}{\delta s}, \pounds_vs \right\rangle 
& = - d_{\frac{\delta \ell}{\delta u}} \left( 1, j_{\frac{\delta l}{\delta u}}, v \right), \qquad \forall\; v\in\mathfrak{g}_0\\
\left\langle w, \mathcal{D}_t s \right\rangle 
& =
d_{\frac{\delta \ell}{\delta u}} \left( \frac{w}{\frac{\delta \ell}{\delta s}}, j_{\frac{\delta l}{\delta u}}, u \right) ,\quad \forall w\in\Lambda^0(\Omega).
\end{align}
\end{subequations}
In strong form, the same equations read
\begin{subequations}
\begin{align}
\frac{\partial }{\partial t}\frac{\delta \ell}{\delta u}
+ \pounds_{u} \frac{\delta \ell}{\delta u} 
& =
\frac{\delta \ell}{\delta a} \diamond a 
+ \frac{\delta \ell}{\delta s} \diamond s
- \mathfrak{d} j_{\frac{\delta \ell}{\delta u}}, \\
\mathcal{D}_t s
& =
\frac{w}{\frac{\delta \ell}{\delta s}} \nabla^g u \ \dot{\wedge} \ j_{\frac{\delta l}{\delta u}}.
\end{align}
\end{subequations}
\begin{remark}
The viscosity function was directly given in Eulerian (i.e. reduced) form. Contrary to the case of differential forms where $d_a = d_A$, its Lagrangian counterpart $d_{\frac{\delta L}{\delta \dot{\varphi}}}$ is not strictly the same. The reason comes from the dependence on a metric, meaning that the metric also needs to be pullbacked.
\end{remark}

\paragraph{General flux closure.} Including cross-effects with differential forms $\{x^\alpha\}_{\alpha \in \mathcal{A}}$, the most general flux closures are
\begin{subequations}
\begin{align}
j_{\frac{\delta \ell}{\delta u}} & = \mathcal{K}_{uu} \nabla^g u + \sum_{\beta \in \mathcal{A}} \mathcal{K}_{u \beta} x^\beta, \\
j_\alpha & = \sum_{\beta \in \mathcal{A}} \mathcal{K}_{\alpha \beta} x^\beta + \mathcal{K}_{\alpha u} \nabla^g u, & \forall \;\alpha \in \mathcal{A}.
\end{align}
\end{subequations}
The new linear operators
\begin{equation}
\mathcal{K}_{uu} : W^* \rightarrow W, \quad \mathcal{K}_{\alpha u} : W^* \rightarrow \Lambda^{n-\text{deg}(x^\alpha)}(\Omega), \quad \mathcal{K}_{u\alpha} : \Lambda^{n-\text{deg}(x^\alpha)}(\Omega) \rightarrow W,
\end{equation}
must satisfy Onsager's principle. Since $\nabla^g u$ is kept invariant by time reversal, it reads
\begin{subequations}
\begin{align}
\mathcal{K}_{uu} & = \mathcal{K}_{uu}^T, \\
\label{eq:Onsager_alpha_u}
\mathcal{K}_{\alpha u} & = \pm \mathcal{K}_{u \alpha}^T, & \forall \;\alpha \in \mathcal{A}_{\pm}.
\end{align}
\end{subequations}
In \eqref{eq:Onsager_alpha_u}, the transpose is to be understood as $x \wedge (\mathcal{K}_{\alpha u} \mathcal{S}) = \mathcal{S} \ \dot{\wedge} \ (\mathcal{K}_{\alpha u}^T x)$, for all $x, \mathcal{S}$.

\paragraph{Isotropic viscous systems.} For isotropic applications, it is first necessary to write the representation of $O(n)$ on $W^*_x = T_x \Omega \otimes T^*_x \Omega$, and then decompose it in irreducible subspaces. The canonical action is $\cdot_1 \triangleright_2$, which is the composition of the action \eqref{eq:O(n)_vectors} on the first component and the action \eqref{eq:O(n)_forms_1} on the second component. The action on $W_x = T^*_x \Omega \otimes \Lambda^{n-1} T^*_x \Omega$ that preserves the density-valued dual pairing is then $\triangleright_1 \triangleleft_2$
\begin{equation}
(R \cdot_1 \triangleright_2 \mathcal{S}) \ \dot{\wedge} \ (R \triangleright_1 \triangleleft_2 \mathcal{T}) = \mathcal{S} \ \dot{\wedge} \ \mathcal{T}, \quad \forall\; (\mathcal{S}, \mathcal{T}) \in W^*_x \times W_x.
\end{equation}
The representation splits into three pairwise non-isomorphic irreducible subrepresentations of the real type $W^*_x = W^h \oplus W^0 \oplus W^a$ \cite{Zee2016,Groot1962}, where $W^h$ corresponds to the homogeneous part of the tensor, $W^0$ the trace-free symmetric part, and $W^a$ the skew-symmetric part. For $\mathcal{S} \in W^*_{x, \cdot_1 \triangleright_2}$, the decomposition reads
\begin{equation}
\mathcal{S} = \frac{\text{tr}(\mathcal{S})}{n} \mathbb{I} + \left( \frac{1}{2}\left( \mathcal{S} + \mathcal{S}^T \right) - \frac{\text{tr}(\mathcal{S})}{n} \mathbb{I} \right) + \frac{1}{2}\left( \mathcal{S} - \mathcal{S}^T \right),
\end{equation}
where $\mathcal{S}^T$ is the metric adjoint of $\mathcal{S}$ when seen as an endomorphism of the tangent space. In order to find all intertwining operators $\mathcal{K}_{uu} \in \text{Hom}_{O(n)}(W^*_{x}, W_{x})$, first define $K_{uu} = \sharp_1 \star^{-1}_2 \mathcal{K}_{uu}$ (or $\mathcal{K}_{uu} = \flat_1 \star_2 K_{uu}$). Then, for all $R \in O(n)$,
\begin{subequations}
\begin{align}
\mathcal{K}_{uu} R \cdot_1 \triangleright_2 = R \triangleright_1 \triangleleft_2 \mathcal{K}_{uu}
& \Longleftrightarrow
(\flat_1 \star_2 K_{uu}) R \cdot_1 \triangleright_2 = R \triangleright_1 \triangleleft_2 (\flat_1 \star_2 K_{uu}) \\
& \Longleftrightarrow
\flat_1 \star_2 (K_{uu} R \cdot_1 \triangleright_2) = \flat_1 \star_2 (R \cdot_1 \triangleright_2 K_{uu}) \\
& \Longleftrightarrow
K_{uu} R \cdot_1 \triangleright_2 = R \cdot_1 \triangleright_2 K_{uu},
\end{align}
\end{subequations}
so that
\begin{equation}
\mathcal{K}_{uu} \in \text{Hom}_{O(n)}(W^*_{x}, W_{x}) 
\Longleftrightarrow
K_{uu} \in \text{Hom}_{O(n)}(W^*_{x}, W^*_{x}).
\end{equation}
Following \eqref{eq:classification_intertwining_operators}, $K_{uu}$ is shown to act as an homothety on each irreducible subrepresentation. Thus, there exists $\kappa^h_{uu}, \kappa^0_{uu}, \kappa^a_{uu} \in \mathbb{R}$
\begin{equation}
\label{eq:complete_decomposition_viscosity}
K_{uu} \mathcal{S} = \kappa^h_{uu} \frac{\text{tr}(\mathcal{S})}{n} \mathbb{I} + \kappa^0_{uu} \left( \frac{1}{2}\left( \mathcal{S} + \mathcal{S}^T \right) - \frac{\text{tr}(\mathcal{S})}{n} \mathbb{I} \right) + \kappa^a_{uu} \frac{1}{2}\left( \mathcal{S} - \mathcal{S}^T \right).
\end{equation}
Compliance with the second principle of thermodynamics leads to $\kappa^h_{uu} \geq 0$, $\kappa^0_{uu} \geq 0$, and $\kappa_{uu}^a \geq 0$. The bulk viscosity coefficient $\kappa^h_{uu}$ measures dissipation induced by compression and extension of the fluid. $\kappa^0_{uu}$ is the so-called standard viscosity coefficient and measures dissipation induced by shear. As for coefficient $\kappa^a_{uu}$, it measure dissipation induced by pure rotations of the fluid. For most applications, this is not possible and $\kappa^a_{uu} = 0$; this also ensures that viscosity does not generate an infinite torque and allows for conservation of the angular momentum \cite{Landau1987}.

\begin{remark}\label{remark_cross_effect}
It turns out that $W^h$ is isomorphic to both $V^0_{\triangleright}$ and $V^n_{\triangleleft}$. Likewise $W^a$ is isomorphic to both $V^2_{\triangleright}$ and $V^{n-2}_{\triangleleft}$. The isomorphisms are respectively given by $\lambda \mathbb{I} \leftrightarrow \lambda \leftrightarrow \lambda \text{d} x$ and $\text{d} x^i \otimes \partial_{x^j} - \text{d} x^j \otimes \partial_{x^i} \leftrightarrow \text{d} x^i \wedge \text{d} x^j \leftrightarrow \star (\text{d} x^i \wedge \text{d} x^j)$ (for an orthonormal basis). It is thus possible to have non-trivial cross-effects $\mathcal{K}_{u \alpha}/\mathcal{K}_{\alpha u}$ between viscosity and other thermodynamic forces.
\end{remark}

\section{A case study: dissipative two-species magnetohydrodynamics}
\label{sec:case_study}

In this section, the previous formalism is illustrated with the equations of dissipative magnetohydrodynamics (MHD) with two species. It is shown that the classical equations on $\Omega \subset \mathbb{R}^3$ are recovered, and that numerous dissipation sources can be painlessly incorporated. Such sources include: viscosity, heat diffusion, resistivity, mass diffusion, chemical reactions and possible cross-effects. 

\paragraph{Physical quantities.} The MHD equations describe the flow of a electrically conducting fluid and its interaction with the magnetic induction field. Here the fluid is assumed to consist of two distinct species. In the reduced case, the state of the system is then described by a velocity $u \in \mathfrak{g}_0$, an entropy $s \in \Lambda^3(\Omega)$, and three auxiliary variables: the particle number density of the two species $\nu_1, \nu_2 \in \Lambda^3(\Omega)$ and the magnetic induction field $\beta \in \Lambda^2(\Omega)$. The latter is also assumed to satisfy the essential boundary condition $i^*_{\partial \Omega \rightarrow \Omega} \beta = 0$. Given the metric on $\Omega$ induced by the usual scalar product on $\mathbb{R}^3$, the reduced MHD Lagrangian is written 
\begin{equation}
\ell(u, \nu_1, \nu_2, \beta, s) = \int_\Omega \mathfrak{l}(u, \nu_1, \nu_2, \beta, s) = \int_\Omega \frac{1}{2} (u \cdot u)\varrho - \varepsilon(\nu_1, \nu_2, s) - \frac{1}{2\mu_0} \beta \wedge (\star \beta),
\end{equation}
where $\varepsilon \in \Lambda^3(\Omega)$ is the internal energy density and $\varrho = M_1 \nu_1 + M_2 \nu_2 \in \Lambda^3(\Omega)$ is the total mass density. The constant $M_1$ and $M_2$ are the molar mass of the two species. The two species are assumed to be involved in the chemical reaction $1 \leftrightarrow k \times 2$ where $k = \frac{M_1}{M_2}$ is a positive integer.

The geometric variables are related to the usual physical quantities in the following way. The scalar mass density $\rho$ is simply $\rho = \star \varrho$, or, given a metric and a positively oriented orthonormal basis $(e_1, e_2, e_3)$,
\begin{equation}
\varrho = \rho \text{d}x^1 \wedge \text{d}x^2 \wedge \text{d}x^3 = \rho \text{d} x.
\end{equation}
Likewise, the scalar particle number density, the scalar entropy density and scalar internal energy are respectively $n_i = \star \nu_i$, $\mathsf{s} = \star s$ and $e = \star \varepsilon$. The magnetic induction field $\beta$ is related to its classical vector representation $B$ through $\beta = \star (B^{\flat})$ (or $B = (\star \beta)^{\#}$). In the same coordinate system, it reads
\begin{equation}
\beta 
= B^3 \text{d} x^1 \wedge \text{d}x^2
+ B^1 \text{d} x^2 \wedge \text{d}x^3
+ B^2 \text{d} x^3 \wedge \text{d}x^1.
\end{equation}
The dual variables are given as functions of the chemical potentials $\mu_i$, temperature $T$ and magnetic field $H$
\begin{subequations}
\begin{align}
\tilde{\mu}_i & = - \frac{\delta \ell}{\delta \nu_i} = \frac{\partial e}{\partial n_i} - \frac{1}{2} (u \cdot u) M_i = \mu_i - \frac{1}{2} (u \cdot u) M_i, \\
T & = - \frac{\delta \ell}{\delta s} = \frac{\partial e}{\partial \mathsf{s}}, \\
H & = - \left( \star \frac{\delta \ell}{\delta \beta} \right)^\# = \frac{1}{\mu_0} B
\end{align}
\end{subequations}
Additionally, the pressure is defined as $p = \mu_1 n_1 + \mu_2 n_2 + T\mathsf{s} - e$. As for the momentum $m$, it is defined weakly by
\begin{equation}
\left\langle \frac{\delta \ell}{\delta u}, v \right\rangle = \int_\Omega m \cdot v \text{d} x,
\end{equation}
for all vector fields $v \in \mathfrak{g}_0$. For the present Lagrangian, it is simply $m = \rho u$. 

\paragraph{Thermodynamical affinities and fluxes.} Thermodynamical affinities for heat transfer, resistivity, mass diffusion, chemical reaction and viscosity are respectively
\begin{equation}
\text{d} \frac{\delta \ell}{\delta s} \in \Lambda^1(\Omega), 
\quad 
\text{d} \frac{\delta \ell}{\delta \beta} \in \Lambda^2(\Omega),
\quad
\text{d} (-\mu) \in \Lambda^1(\Omega),
\quad 
\text{and}
\quad
-\mu \in \Lambda^0(\Omega),
\end{equation}
where
\begin{equation}\label{mu_expression}
\mu = \frac{1}{M_2} \frac{\delta \ell}{\delta \nu_2} - \frac{1}{M_1} \frac{\delta \ell}{\delta \nu_1} = \frac{1}{M_1} \tilde{\mu}_1  - \frac{1}{M_2} \tilde{\mu}_2 = \frac{1}{M_1} \mu_1  - \frac{1}{M_2} \mu_2.
\end{equation}
The definition of $\mu$ is motivated by mass conservation. The associated thermodynamical fluxes and their physical representation are written
\begin{subequations}
\begin{align}
j_s & \in \Lambda^2(\Omega), & j_\mathsf{s} = (\star j_s)^\sharp & \in \mathfrak{X}(\Omega), \\
j_\beta & \in \Lambda^1(\Omega), & j_B = (j_\beta)^\sharp & \in \mathfrak{X}(\Omega), \\
j^c_\nu & \in \Lambda^2(\Omega), & j^c_n = (\star j^c_\nu)^\sharp & \in \mathfrak{X}(\Omega), \\
j^d_\nu & \in \Lambda^3(\Omega), & j^d_n = \star j^d_\nu & \in \Lambda^0(\Omega).
\end{align}
\end{subequations}
They are chosen so as to satisfy
\begin{subequations}
\begin{align}
\text{d} \frac{\delta \ell}{\delta s} \wedge j_s & = - (j_\mathsf{s} \cdot \nabla T) \text{d} x, &
\text{d} \frac{\delta \ell}{\delta \beta} \wedge j_\beta & = - (j_B \cdot \nabla \times H) \text{d} x, \\
\text{d} (-\mu) \wedge j^d_\nu & = - (j^d_n \cdot \nabla \mu) \text{d} x, &
j^d_\nu \wedge (-\mu) & = - \mu j^d_n \text{d} x.
\end{align}
\end{subequations}
As for the viscosity, the thermodynamic affinity is $\nabla^g u$ and the thermodynamical flux is $j_{\frac{\delta \ell}{\delta u}}$. Their respective physical representation $\nabla u$ and $\sigma$ are given by
\begin{subequations}
\begin{align}
\nabla^g u & \in \Gamma(T \Omega \otimes T^* \Omega), & \nabla u = \sharp_2 \nabla^g u & \in \Gamma(T \Omega \otimes T \Omega), \\
j_{\frac{\delta \ell}{\delta u}} & \in \Gamma(T^* \Omega \otimes \Lambda^{n-1} T^* \Omega) & \sigma = \sharp_1 \sharp_2 \star_2 j_{\frac{\delta \ell}{\delta u}} & \in \Gamma(T \Omega \otimes T \Omega),
\end{align}
\end{subequations}
so that
\begin{equation}
\nabla^g u \ \dot{\wedge} \ j_{\frac{\delta \ell}{\delta u}} = (\sigma : \nabla u) \text{d} x.
\end{equation}
where $:$ is the euclidean metric applied to the two tensors. Adapting the variation principle \ref{vp:4}, the reduced variational principle for two-species magnetohydrodynamics is now presented.
\begin{varia}
Given a reduced Lagrangian $\ell: \mathfrak{g}_0 \times \Lambda^3(\Omega) \times \Lambda^3(\Omega) \times \Lambda^3(\Omega) \times \Lambda^2(\Omega) \times \Lambda^3(\Omega) \rightarrow \mathbb{R}$, the variational problem consists in finding $u: [0, T] \rightarrow \mathfrak{g}_0$, $\nu_i: [0, T] \rightarrow \Lambda^3(\Omega)$, $\chi_i: [0, T] \rightarrow \Lambda^0(\Omega)$, $i \in \{1, 2\}$, $\beta: [0, T] \rightarrow \Lambda^2(\Omega)$, $\theta: [0, T] \rightarrow \Lambda^1(\Omega)$, $s: [0, T] \rightarrow \Lambda^3 (\Omega)$, $\varsigma: [0, T] \rightarrow \Lambda^3 (\Omega)$, and $\gamma: [0, T] \rightarrow \Lambda^0 (\Omega)$ such that
\begin{multline}
\delta \int_0^T \ell(u(t), \nu_1(t), \nu_2(t), \beta(t), s(t)) 
+ \left\langle \mathcal{D}_t \chi_1(t), \nu_1(t) \right\rangle
+ \left\langle \mathcal{D}_t \chi_2(t), \nu_2(t) \right\rangle \\
+ \left\langle \mathcal{D}_t \theta(t), \beta(t) \right\rangle
+ \left\langle \mathcal{D}_t \gamma (t), s(t) - \varsigma(t) \right\rangle {\rm d} t = 0
\end{multline}
with $\delta u = \dot{v} + [u,v]$, $v \in \mathfrak{g}_0$, and the phenomenological and variational constraints
\begin{subequations}
\begin{align}
\notag \left\langle w, \mathcal{D}_t \varsigma \right\rangle 
& =
d_{\frac{\delta \ell}{\delta u}} \left( \frac{w}{\frac{\delta \ell}{\delta s}}, j_\frac{\delta \ell}{\delta u}, u \right)
+ d^c \left( \frac{w}{\frac{\delta \ell}{\delta s}}, j_s, \mathcal{D}_t \gamma \right)
+ d^d \left( \frac{w}{\frac{\delta \ell}{\delta s}}, j^d_{\nu}, \frac{1}{M_1} \mathcal{D}_t \chi_1 - \frac{1}{M_2} \mathcal{D}_t \chi_2 \right) \\
& \hspace{1cm} 
+ d^c \left( \frac{w}{\frac{\delta \ell}{\delta s}}, j^c_{\nu}, \frac{1}{M_1} \mathcal{D}_t \chi_1 - \frac{1}{M_2} \mathcal{D}_t \chi_2 \right)
+ d^c \left( \frac{w}{\frac{\delta \ell}{\delta s}}, j_\beta, \mathcal{D}_t \theta \right),\quad \forall w \in \Lambda^0(\Omega) \\
\notag \left\langle w, \mathcal{D}_\delta \varsigma \right\rangle
& = 
d_{\frac{\delta \ell}{\delta u}} \left( \frac{w}{\frac{\delta \ell}{\delta s}}, j_\frac{\delta \ell}{\delta u}, v \right)
+ d^c \left( \frac{w}{\frac{\delta \ell}{\delta s}}, j_s, \mathcal{D}_\delta \gamma \right)
+ d^d \left( \frac{w}{\frac{\delta \ell}{\delta s}}, j^d_{\nu}, \frac{1}{M_1} \mathcal{D}_\delta \chi_1 - \frac{1}{M_2} \mathcal{D}_\delta \chi_2 \right) \\
& \hspace{1cm} 
+ d^c \left( \frac{w}{\frac{\delta \ell}{\delta s}}, j^c_{\nu}, \frac{1}{M_1} \mathcal{D}_\delta \chi_1 - \frac{1}{M_2} \mathcal{D}_\delta \chi_2 \right)
+ d^c \left( \frac{w}{\frac{\delta \ell}{\delta s}}, j_\beta, \mathcal{D}_\delta \theta \right),\quad \forall w \in \Lambda^0(\Omega).
\end{align}
\end{subequations}
\end{varia}

\paragraph{Resulting equations.} 
Following the approach developed in the previous sections, the equations resulting from the variational principle are
\begin{subequations}
\begin{align}
\frac{\partial}{\partial t} \frac{\delta \ell}{\delta u} + \pounds_u \frac{\delta \ell}{\delta u} & = \sum_{x \in \{\nu_1, \nu_2, \beta, s \}} \frac{\delta \ell}{\delta x} \diamond x + \mathfrak{d} j_{\frac{\delta \ell}{\delta u}}, \\
-\frac{\delta \ell}{\delta s} \mathcal{D}_t s & =
\nabla^g u \ \dot{\wedge} \ j_{\frac{\delta \ell}{\delta u}} 
+
\text{d} \left( \frac{\delta \ell}{\delta s} \wedge j_s \right)
+
(-\mu) \wedge j^d_{\nu} 
+ 
\text{d} (-\mu) \wedge j^c_{\nu}
+ 
\text{d} \frac{\delta \ell}{\delta \beta} \wedge j_\beta, \\
\mathcal{D}_t \nu_1 & = +\frac{1}{M_1} \left( j^d_\nu + \text{d} j^c_{\nu} \right), \\
\mathcal{D}_t \nu_2 & = -\frac{1}{M_2} \left( j^d_\nu + \text{d} j^c_{\nu} \right), \\
\mathcal{D}_t \beta & = \text{d} j_{\beta},
\end{align}
\end{subequations}
where we recall that $\mu$ is given in \eqref{mu_expression}.
The equations are supplemented with the boundary conditions
\begin{subequations}
\label{eq:MHD_bd}
\begin{align}
i^*_{\partial \Omega \rightarrow \Omega} \beta & = 0, &
i^*_{\partial \Omega \rightarrow \Omega} j_s & = 0, \\
i^*_{\partial \Omega \rightarrow \Omega} j^c_{\nu} & = 0, &
i^*_{\partial \Omega \rightarrow \Omega} j_\beta & = 0,
\end{align}
\end{subequations}
also emerging from the variational principle. Using the usual representation of the physical quantities, the system of equation becomes
\begin{subequations}
\label{eq:full_MHD}
\begin{align}
\label{eq:u_usual_form}
\partial_t (\rho u) + \nabla \left( \rho u \otimes u \right) & = - \nabla p + \left[(\nabla \times H) \times B + (\nabla \cdot B) H\right] + \nabla \cdot \sigma, \\
\notag
T \left( \partial_t \mathsf{s} + \nabla \cdot (\mathsf{s} u) \right) & = \sigma : \nabla u - \nabla \cdot (T j_\mathsf{s}) - j^d_{n} \mu - j^c_{n} \cdot \nabla \mu - j_B \cdot \nabla \times H, \\
\partial_t n_1 + \nabla \cdot (n_1 u) & = + \frac{1}{M_1} \left( j^d_n + \nabla \cdot j^c_n \right), \\
\partial_t n_2 + \nabla \cdot (n_2 u) & = - \frac{1}{M_2} \left(  j^d_n + \nabla \cdot j^c_n \right), \\
\label{eq:B_usual_form}
\partial_t B + \nabla \times ( B \times u) + (\nabla \cdot B) u & = \nabla \times j_B.
\end{align}
\end{subequations}
As for the boundary conditions \eqref{eq:MHD_bd}, they can now be expressed as a function of the outward normal $n$ at the boundary and read
\begin{subequations}
\begin{align}
B \cdot n & = 0, & j_\mathsf{s} \cdot n & = 0, \\
j^c_n \cdot n & = 0, & j_B \times n & = 0.
\end{align}
\end{subequations}

\begin{remark}
MHD equations satisfy the additional solenoidal condition $\nabla \cdot B = 0$ (i.e. $\text{d} \beta = 0$). Once it is enforced, equations \eqref{eq:u_usual_form}-\eqref{eq:B_usual_form} can be simplified into
\begin{subequations}
\begin{align}
\partial_t (\rho u) + \nabla \left( \rho u \otimes u \right) & = - \nabla p + (\nabla \times H) \times B + \nabla \cdot \sigma, \\
\partial_t B + \nabla \times ( B \times u) & = \nabla \times j_B.
\end{align}
\end{subequations}
While the latter expressions are more common, the former \eqref{eq:u_usual_form} and \eqref{eq:B_usual_form} are essential to show momentum conservation and symmetrization of the system of equations \cite{Godunov2025}. This thus emphasizes the strength and legitimacy of the geometric formulation.
\end{remark}

\paragraph{Conservation laws.} 
Conservation of energy, conservation of mass and conservation of the magnetic flux can be derived and respectively yield
\begin{subequations}
\label{eq:conservation_laws}
\begin{align}
\frac{\text{d}}{\text{d} t} \int_\Omega (u \cdot u) \varrho + \varepsilon(n_1, n_2,s) + \frac{1}{2\mu_0} \beta \wedge \star \beta = 0, \\
\frac{\text{d}}{\text{d} t} \int_{\mathcal{V}(t)} \varrho = 0, \\
\frac{\text{d}}{\text{d} t} \int_{\mathcal{S}(t)} i^*_{\mathcal{S}(t) \rightarrow \Omega} \beta = \int_{\partial \mathcal{S}(t)} i^*_{\partial \mathcal{S}(t) \rightarrow \Omega} j_\beta,
\end{align}
\end{subequations}
for every volume $\mathcal{V}(t)$ and surface $\mathcal{S}(t)$ moving with the flow. It is also possible to derive a similar conservation equation for the exchanged entropy $s-\varsigma$. With the usual physical representation of the quantities, \eqref{eq:conservation_laws} read
\begin{subequations}
\begin{align}
\frac{\text{d}}{\text{d} t} \int_\Omega \rho \| u \|^2 + e(n_1, n_2, \mathsf{s}) + \frac{1}{2\mu_0} \| B \|^2 \text{d} x = 0, \\
\frac{\text{d}}{\text{d} t} \int_{\mathcal{V}(t)} \rho \text{d} x = 0, \\
\frac{\text{d}}{\text{d} t} \int_{\mathcal{S}(t)} B \cdot \text{d} S = \int_{\partial \mathcal{S}(t)} j_{B} \cdot \text{d} l.
\end{align}
\end{subequations}
The last equation recovers Alfv\'en's theorem, extended to resistive flows. Additionally, energy conservation can be given the following local form
\begin{equation}
\mathcal{D}_t \mathfrak{e} = \text{d} \left[ \sum_{\phi \in \{ \nu_1, \nu_2, \beta, s \}} (-1)^{n-k_\phi} \frac{\partial \mathfrak{l}}{\partial \phi} \wedge i_u \phi - i_u \mathfrak{l} + \frac{\partial \mathfrak{l}}{\partial s} \wedge j_s + (-\mu) \wedge j^c_\nu + \frac{\partial \mathfrak{l}}{\partial \beta} \wedge j_\beta + u \cdot_1 j_{\frac{\delta \ell}{\delta u}} \right].
\end{equation}
With the usual physical notations, the evolution equation of $e = \star \mathfrak{e}$ reads
\begin{equation}
\partial_t e + \nabla \cdot (eu) = \nabla \cdot \left[ -pu + (B \times u) \times H - T j_\mathsf{s} -\mu j^c_n + j_B \times H + \sigma \cdot u \right].
\end{equation} 

\paragraph{Isotropic flux closures and entropy.} 
In general, coupling can be written between every pair of thermodynamic affinities. The temperature and the chemical potentials are invariant by time reversal symmetry, while the magnetic field and deformation gradient are not. For isotropic applications, one needs to look at how they transform under the action of the orthogonal group. In physics, the temperature and chemical potential gradients are polar vectors so $O(3)$ acts on their 1-form counterpart with $\triangleright$. The rotational of $H$ is also a polar vector as $H$ is a pseudo-vector and the rotational operator transforms pseudo-vectors into polar vectors; $O(3)$ then must act on its 2-form counterpart with $\triangleleft$. Since $V^1_{\triangleright}$ and $V^2_{\triangleleft}$ are isomorphic as representations, isotropic cross-effects can be written between all of them and they read
\begin{subequations}
\begin{align}
j^d_\nu & = - \star \kappa_\nu \mu, \\
j_s & =
\kappa_{ss} \star \text{d} \frac{\delta \ell}{\delta s}
+
\kappa_{\nu s} \star \text{d} (-\mu)
- 
\kappa_{\beta s} \phantom{{} \star {}} \text{d} \frac{\delta \ell}{\delta \beta}, \\
j^c_\nu & =
\kappa_{\nu s} \star \text{d} \frac{\delta \ell}{\delta s}
+
\kappa_{\nu \nu} \star \text{d} (-\mu)
- 
\kappa_{\beta \nu} \phantom{{} \star {}} \text{d} \frac{\delta \ell}{\delta \beta}, \\
j_\beta & =  
\kappa_{\beta s} \phantom{{} \star {}} \text{d} \frac{\delta \ell}{\delta s}
+
\kappa_{\beta \nu} \phantom{{} \star {}} \text{d} (-\mu)
+ 
\kappa_{\beta \beta} \star \text{d} \frac{\delta \ell}{\delta \beta}.
\end{align}
\end{subequations}
When translating to the usual physical notations, $\kappa_{\nu \nu}$ is replaced by $\kappa_{nn}$ and likewise for the others. The value is the same but the new notations emphasizes the change of variables. Eventually,
\begin{subequations}
\begin{align}
j^d_n & = - \kappa_n \mu, \\
\begin{pmatrix}
j_{\mathsf{s}} \\
j^c_n \\
j_{B}
\end{pmatrix}
& =
\begin{pmatrix}
\kappa_{\mathsf{s}\mathsf{s}} & \kappa_{n \mathsf{s}} & -\kappa_{B \mathsf{s}} \\
\kappa_{n \mathsf{s}} & \kappa_{n n} & -\kappa_{B n} \\
\kappa_{B \mathsf{s}} & \kappa_{B n} & \kappa_{B B}
\end{pmatrix}
\begin{pmatrix}
- \nabla T \\
- \nabla \mu \\
- \nabla \times H
\end{pmatrix}.
\end{align}
\end{subequations}
Diagonal coefficients $\kappa_{\mathsf{s} \mathsf{s}}, \kappa_{n n}$ and $\kappa_{B B}$ are respectively related to heat transfer, particular diffusion and resistivity. Coefficients $\kappa_{\mathsf{s} n}$, $\kappa_{\mathsf{s} B}$ and $\kappa_{n B}$ respectively describe the cross-effect of thermo-diffusion (Soret-Dufour), thermo-electric (Seebeck-Peltier) and electro-diffusion. Finally, in order for the entropy production to be positive, the symmetric part of the global matrix must be positive-definite, or, equivalently
\begin{subequations}
\begin{align}
\kappa_{\mathsf{s} \mathsf{s}} & \geq 0, & \kappa_{nn} & \geq 0, \\
\kappa_{BB} & \geq 0, & \kappa_{\mathsf{s} \mathsf{s}} \kappa_{nn} - \kappa_{\mathsf{s} n}^2 & \geq 0.
\end{align}
\end{subequations}

\begin{remark}
The present model naturally extends to an arbitrary number of species with chemical reactions involving more reagents. Each chemical reaction is then described by its own thermodynamical affinity which is a linear combination $\sum_i \lambda_i \tilde{\mu}_i$. Because of mass conservation, the coefficients must satisfy $\sum_{i} \lambda_i M_i = 0$ which then allows to write $\sum_i \lambda_i \tilde{\mu}_i = \sum_i \lambda_i \mu_i$. Cross-terms can also be written between the compressive part of the viscosity (i.e. $\propto \nabla \cdot u$) and thermodynamics affinities associated to chemical reactions since they are all isomorphic (trivial) representations under the action of $O(3)$. If the skew-symmetric part of the deformation tensor were to be added as in \eqref{eq:complete_decomposition_viscosity}, no cross-effect could be written as there is no affinity belonging to $V^1_{\triangleleft}$ or $V^2_{\triangleright}$. Finally, the expression of the Lagrangian can account for additional potentials such as gravity or surface tension.
\end{remark}

\section{Conclusion}

An extended Hamilton's principle was presented for dissipative fluid equations, where variables are written as differential forms. The geometric nature of the equations is emphasized so they can directly be written on an arbitrary oriented manifolds. Dissipation is included in the variational principle in a way that is automatically consistent with the first law of thermodynamics (i.e. energy conservation), regardless of the chosen thermodynamic fluxes. Closures for these fluxes are then investigated in order to comply with the second law of thermodynamics, Onsager's reciprocity principle, as well as with symmetries of the system through Curie's principle. Finally, the formulation is illustrated with the equations of multi-species MHD, confirming the effectiveness of the method and its ability to recover physically relevant equations.

The present formalism relies on exterior calculus and differential forms. Extending it to tensor-valued fields is an important direction as it would allow to systematically treat generalized viscous-like terms on arbitrary Riemannian manifolds. The method could also be extended to multi-velocity and multi-entropy systems; such systems display much more intricate dissipation processes. Finally, this work may serve as a foundation for structure-preserving discretizations. Mimicking the geometric formulation at the discrete level would naturally produce schemes that are compatible with thermodynamics, a feature which is essential for both accuracy and stability of simulations.

\paragraph{Ackowledgement.}
We are grateful to Xavier Garbet for fruitful discussions.
Fran\c{c}ois Gay-Balmaz was supported by a startup grant from Nanyang Technological University.
Fran\c{c}ois Gay-Balmaz and Bastien Manach-P\'erennou were supported by the National Research
Foundation Singapore (NRF) core funding ‘Fusion Science for Clean Energy’. 

\bibliographystyle{plain}
\bibliography{bib.bib}

\end{document}